\documentclass[aip,amsmath,amssymb,reprint]{revtex4-2}
\usepackage{epsfig}                                                                 
\usepackage{dcolumn}                                                                
\usepackage{bm}                                                                     
\usepackage{xcolor}
\usepackage{tabularx}

\usepackage[pdfstartview=FitH,bookmarksopen=true,bookmarksopenlevel=1]{hyperref}    

\begin{document}
\title{Does the van der Waals force play a part in evaporation?}
\author{E. S. Benilov}
 \email[Email address: ]{Eugene.Benilov@ul.ie}
 \homepage[\newline Homepage: ]{https://eugene.benilov.com/}
 \affiliation{Department of Mathematics and Statistics, University of Limerick, Limerick V94~T9PX, Ireland}

\begin{abstract}
It is argued that the van der Waals force exerted by the liquid and vapor/air
on the molecules escaping from one phase into the other strongly affects the
characteristics of evaporation. This is shown using two distinct descriptions
of the van der Waals force: the Vlasov and diffuse-interface models, each of
which is applied to two distinct settings: a liquid evaporating into its vapor
and a liquid evaporating into air (in all cases, the vapor-to-liquid density
ratio is small). For the former setting, the results are consistent with the
Hertz--Knudsen Law (HKL), but the evaporation/condensation probability is very
small (in the classical HKL, it is order one). For the latter setting, the
dependence of the evaporation rate on the difference between the saturated
vapor pressure and its actual value is shown to be nonlinear (whereas the
classical HKL predicts a linear dependence). The difference between the two
settings indicates that the van der Waals force exerted by the air strongly
affects evaporation (contrary to the general assumption that the ambient gas
is unimportant). Finally, the diffuse interface model is shown to be
inapplicable in a narrow region at the outskirts of the interface -- as a
result, it noticeably underestimates the evaporative flux by comparison with
the (more accurate) Vlasov model.

\end{abstract}
\maketitle

\section{Introduction\label{sec1}}

Evaporation is fundamental to numerous environmental, biological, and
industrial processes, and the Hertz--Knudsen Law (HKL) is our primary tool for
modeling it. This paper argues that the HKL needs to be modified by taking
into account the effect of the van der Waals force on the evaporative flux.

In its original formulation \cite{Hertz82,Knudsen15}, the HKL was based on an
assumption that the flux of molecules escaping from a liquid into vapor does
not depend on the vapor pressure -- hence, this flux can be calculated as if
the vapor were saturated. Calculating also the flux in the opposite direction
(which does depend on the actual vapor pressure), one can show that the net
evaporative flux is%
\begin{equation}
E=\sqrt{\frac{RT}{2\pi}}\left(  \rho^{(v.sat)}-\rho^{(v)}\right)  ,
\label{1.1}%
\end{equation}
where $\rho^{(v.sat)}$ is the saturated vapor density, $\rho^{(v)}$ the actual
density, $R$ the specific gas constant, and $T$ the temperature (assumed, for
simplicity, to be the same in the liquid and vapor).

Expression (\ref{1.1}) does not involve a single adjustable parameter and,
thus, is unlikely to be accurate for all liquids under all conditions. To make
it more adaptable, one can assume that some of the escaping molecules bounce
back, as do those travelling in the opposite direction. It can be argued
\cite{BondStruchtrup04} that a molecule's probability of evaporation equals
that of condensation, resulting in the following modification of expression
(\ref{1.1}):%
\begin{equation}
E=\theta\,\sqrt{\frac{RT}{2\pi}}\left(  \rho^{(v.sat)}-\rho^{(v)}\right)  ,
\label{1.2}%
\end{equation}
where the evaporation/condensation probability $\theta$ (called also
\textquotedblleft mass adjustment coefficient\textquotedblright) depends on
$T$. The amended version of the HKL still disagrees with some of the available
experiments, and those disagree with each other: for, say, water, the measured
values of $\theta$ vary between $0.01$ and $1$ for the same temperature
\cite{EamesMarrSabir97,MarekStraub01}. There are also several theoretical
models (e.g., \cite{Schrage53,PersadWard16,StruchtrupBeckmannRanaFrezzotti17}%
), but the discord in the experimental results makes it difficult to choose
the most accurate theory.

The present paper is motivated by an observation that none of the existing
models of evaporation accounts for the van der Waals (vdW) force. Yet it is
clearly important: it holds the liquid/vapor interface together (by balancing
the pressure gradient due to the density variation) -- hence, should affect
the molecules passing through the interface.

It can also be argued that the vdW force makes evaporation of a liquid into
its vapor different from evaporation into air. To understand why, note that
the vdW force exerted by the bulk of the liquid pulls the escaping molecules
back and, thus, \emph{impedes} evaporation -- while the vapor and air pull
them forward and, thus, \emph{encourage} evaporation. Since under normal
conditions the vapor and air densities differ by orders of magnitude, the
former exerts a much stronger vdW force than the latter. As a result,
evaporation into air should occur much faster than that into vapor -- and this
is one of the two main conclusions of the present work.

The other one is less intuitive, but still has important physical
implications. As shown for evaporation into air, the vdW force makes the
dependence of the flux $E$ on the density difference $\left(  \rho
^{(v.sat)}-\rho^{(v)}\right)  $ nonlinear, whereas the HKL predicts that
$E\sim\left(  \rho^{(v.sat)}-\rho^{(v)}\right)  $ [see Eq. (\ref{1.2})]. The
two results can be reconciled only if the evaporation/condensation probability
$\theta$ in (\ref{1.2}) depends on $\rho^{(v)}$. Such a dependence could
explain the above-mentioned discord in the measurements of $\theta$ for the
same liquid at the same temperature.

The present paper employs two different descriptions of the vdW force: the
Vlasov model and the diffuse-interface model. The former has been used before
to study interfaces \cite{BongiornoBongiornoScrivenDavis76}, contact lines,
and liquid films \cite{Pismen01,Pismen02,YochelisPismen06} -- but not
evaporation; the latter has been applied to evaporation
\cite{Benilov22a,Benilov23c,Benilov23d}, but its connection to the HKL has not
been properly explored.

The two models of the vdW force will be used in conjunction with the
isothermal Navier--Stokes equations. This simple framework is sufficient for
demonstrating the importance of long-range intermolecular forces for
evaporation \emph{in principle}, and this is the aim of the present paper.

In what follows, Secs. \ref{sec2}--\ref{sec3} examine evaporation of a liquid
into its vapor, Secs. \ref{sec4}--\ref{sec5} examine evaporation into air, and
Sec. \ref{sec6} explains why the (alleged) shortcomings of the HKL have not
been so far observed by the experimentalists and researchers working on
molecular dynamics.

Since the material presented in this paper is associated with a number of
bulky specialized terms, several abbreviations will be used. For future
reference, they are listed in Table \ref{tab1}.

\section{Evaporation of a liquid into its vapor: the formulation\label{sec2}}

\subsection{Thermodynamics\label{sec2.1}}

A model of phase transitions should account for the fluid's thermodynamic
properties. These are described in this subsection, in a brief but
self-consistent manner.

Following \cite{GiovangigliMatuszewski13}, one can fully characterize a fluid
by setting the dependence of its specific (per unit mass) internal energy $e$
and entropy $s$ on the density $\rho$ and temperature $T$. The functions
$e(\rho,T)$ and $s(\rho,T)$ are not arbitrary, but are constrained by the
so-called Gibbs fundamental relation, which can be written in the form%
\begin{equation}
\frac{\partial e}{\partial T}=T\frac{\partial s}{\partial T} \label{2.1}%
\end{equation}
(the equivalence of this equality to the traditional form of the Gibbs
relation is demonstrated in Appendix A of Ref. \cite{Benilov23a}).

Given $e(\rho,T)$ and $s(\rho,T)$, one can find the pressure $p(\rho,T)$ and
chemical potential $G(\rho,T)$ via the formulae%
\begin{align}
p  &  =\rho^{2}\left(  \frac{\partial e}{\partial\rho}-T\frac{\partial
s}{\partial\rho}\right)  ,\label{2.2}\\
G  &  =\frac{\partial\left(  \rho e\right)  }{\partial\rho}-T\frac
{\partial\left(  \rho s\right)  }{\partial\rho}. \label{2.3}%
\end{align}
Eqs. (\ref{2.2})--(\ref{2.3}) imply that%
\begin{equation}
\frac{1}{\rho}\frac{\partial p}{\partial\rho}=\frac{\partial G}{\partial\rho},
\label{2.4}%
\end{equation}
which is the only thermodynamic identity needed in this paper.

\begin{figure}\vspace{5mm}
\begin{center}\includegraphics[width=\columnwidth]{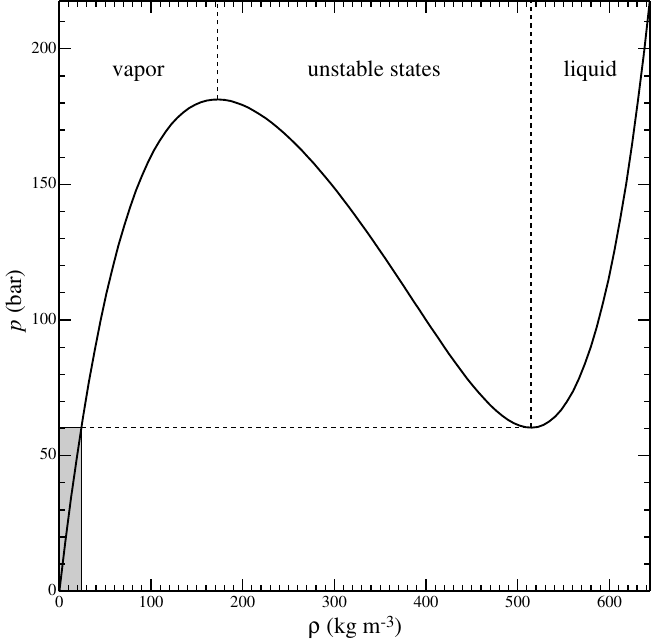}\end{center}
\caption{The pressure $p$ vs. density $\rho$ for the Enskog--Vlasov equation of state (\ref{2.8}) for water at $T=352^{\circ}\mathrm{C}$. The region where the vapor density does not have a match in the liquid region is shaded (it exists only if $T$ is sufficiently high, so that the local minimum of the curve $p$ vs. $\rho$ is above the horizontal axis).}
\label{fig1}
\end{figure}

An example of $p(\rho,T)$ (often referred to as the equation of state) is
shown in Fig. \ref{fig1}. Observe that the dependence of $p$ on $\rho$ is
nonmonotonic: the states between the origin and local maximum are vapor and
those between the local minimum and infinity, liquid. The states between the
minimum and maximum (those with $\partial p/\partial\rho<0$) are unstable. To
understand why, observe that, in this case, a spatially-localized density
\emph{in}crease causes a pressure \emph{de}crease -- the resulting pressure
gradient generates an inward flow -- which causes a further density increase
-- hence, instability.

\begin{table}
\renewcommand{\arraystretch}{1.6}\centering
\begin{ruledtabular}
\caption{Abbreviations used in this paper.\vspace{2mm}}
\begin{tabularx}{\columnwidth}{ll}
Abbreviation & Full form\\[5pt]\hline
HKL & Hertz--Knudsen Law\\
vdW & van der Waals (force, layer)\\
DIM & diffuse-interface model\\
VM & Vlasov model\\
LJ & Lennard-Jones (potential)\\
\end{tabularx}
\label{tab1}
\end{ruledtabular}
\end{table}

The saturated vapor density $\rho^{(v.sat)}$ and the matching liquid density
$\rho^{(l.sat)}$, for a given temperature $T$, satisfy the so-called Maxwell
construction -- i.e.,%
\begin{equation}
p(\rho^{(v.sat)},T)=p(\rho^{(l.sat)},T), \label{2.5}%
\end{equation}%
\begin{equation}
G(\rho^{(v.sat)},T)=G(\rho^{(l.sat)},T). \label{2.6}%
\end{equation}
Conditions (\ref{2.5})--(\ref{2.6}) guarantee that the interface separating
the vapor and liquid is in \emph{mechanical} and \emph{thermodynamic}
equilibrium, respectively.

Due to nonmonotonicity of $p$, the solution of Eqs. (\ref{2.5})--(\ref{2.6})
is not unique. To make it such, require that the vapor and liquid are both
stable%
\[
\frac{\partial p(\rho,T)}{\partial\rho}>0\qquad\text{for}\qquad\rho
=\rho^{(l.sat)},\rho^{(v.sat)}.
\]
The general results reported in this paper are illustrated using the so-called
Enskog--Vlasov fluid model, according to which%
\begin{equation}
e=cT-a\rho,\qquad s=c\ln T-R\ln\rho-R\Theta(\rho), \label{2.7}%
\end{equation}
where $c$ is the specific heat capacity of the fluid under consideration.

The first term in $e$ and the first two in $s$ correspond to ideal gas. The
term involving $a$ describes the contribution of the vdW force to the internal
energy, and the function $\Theta(\rho)$ is the non-ideal contribution to the
entropy. Both $a$ and $\Theta(\rho)$ should be fixed by fitting the fluid's
equation of state to its empiric shape \cite{Benilov23a}; in the present
paper, the values corresponding to water will be used (see Appendix
\ref{appA.1}). The Enskog--Vlasov fluid model is sufficiently accurate (as
shown in Secs. 8.1--2 of Ref. \cite{Benilov23a}), plus it is consistent with
the Vlasov description of the van der Waals force used in this paper.

Substituting the Enskog--Vlasov expressions (\ref{2.7}) into Eqs.
(\ref{2.2})--(\ref{2.3}), one obtains%
\begin{align}
p  &  =T\left[  R\rho+\rho^{2}\frac{\mathrm{d}\Theta(\rho)}{\mathrm{d}\rho
}\right]  -a\rho^{2},\label{2.8}\\
G  &  =T\left[  R\ln\rho+\rho\frac{\mathrm{d}\Theta(\rho)}{\mathrm{d}\rho
}+\Theta(\rho)\right]  -2a\rho+\cdots, \label{2.9}%
\end{align}
where $\cdots$ hides the terms in the chemical potential depending only on $T$
(they cancel out from identity (\ref{2.4}), the Maxwell construction, and all
the equations to come).

In the low-density limit, expressions (\ref{2.8})--(\ref{2.9}) yield%
\begin{equation}
p\sim RT\rho,\qquad G\sim RT\ln\rho\qquad\text{as}\qquad\rho\rightarrow0.
\label{2.10}%
\end{equation}
These asymptotics describe an ideal gas and, thus, are not specific to the
Enskog--Vlasov fluid model.

The first term in expression (\ref{2.8}) will be referred to as the
\emph{thermal} pressure; denoting it by $\hat{p}$, one obtains%
\begin{equation}
\hat{p}=p+a\rho^{2}. \label{2.11}%
\end{equation}
A similar equality inter-relates the thermal and full chemical potentials,%
\begin{equation}
\hat{G}=G+2a\rho. \label{2.12}%
\end{equation}
Eqs. (\ref{2.11})--(\ref{2.12}) and (\ref{2.4}) imply that%
\begin{equation}
\frac{1}{\rho}\frac{\partial\hat{p}}{\partial\rho}=\frac{\partial\hat{G}%
}{\partial\rho}. \label{2.13}%
\end{equation}
Finally, the low-density asymptotics of the thermal chemical potential
coincides with that of the full chemical potential,%
\begin{equation}
\hat{G}\sim RT\ln\rho\qquad\text{as}\qquad\rho\rightarrow0, \label{2.14}%
\end{equation}
whereas the asymptotic of $\hat{p}$ will not be needed.

\subsection{The low-temperature limit\label{sec2.2}}

Assume that the nondimensional temperature $T_{nd}$ is small,%
\begin{equation}
T_{nd}=\frac{RT}{a\rho}\ll1. \label{2.15}%
\end{equation}
This restriction typically applies to all common liquids under normal
conditions; for water between $0^{\circ}\mathrm{C}$ and $100^{\circ}%
\mathrm{C}$, for example, one obtains%
\[
0.065\lesssim T_{nd}\lesssim0.082.
\]
The assumption $T_{nd}\ll1$ implies that the vapor-to-liquid density ratio is
also small -- or, to be precise, the smallness of $T_{nd}$ makes
$\rho^{(v.sat)}/\rho^{(l.sat)}$ \emph{exponentially} small (as shown in Ref.
\cite{Benilov20d} for the generic van der Waals equation of state, but is also
true generally). For water between $0^{\circ}\mathrm{C}$ and $100^{\circ
}\mathrm{C}$, for example, one can use the online calculator
\cite{LindstromMallard97} to obtain%
\[
4.9\times10^{-6}\lesssim\frac{\rho^{(v.sat)}}{\rho^{(l.sat)}}\lesssim
6.2\times10^{-4}.
\]
The assumption $\rho^{(v.sat)}/\rho^{(l.sat)}\ll1$ underlies all results of
this paper.

\subsection{The van der Waals force\label{sec2.3}}

Consider a compressible fluid characterized by its density field
$\rho(\mathbf{r},t)$, where $\mathbf{r}$ is the position vector and $t$, the
time. Introduce also the molecular mass $m$, so that $\rho/m$ is the number density.

Let the potential of the van de Waals force exerted at a point $\mathbf{r}$ by
a molecule located at a point $\mathbf{r}^{\prime}$ be $m^{2}\Phi
(\mathbf{r}-\mathbf{r}^{\prime})$, where the factor $m^{2}$ is introduced for
convenience. If the fluid is isotropic, then%
\[
\Phi(\mathbf{r})=\Phi(r),
\]
where $r=|\mathbf{r|}$. Without loss of generality, one can assume that%
\[
\Phi\rightarrow0\qquad\text{as}\qquad r\rightarrow\infty.
\]
Strictly speaking, $\Phi(r)$ should be determined by examining quantum
interaction of the fluid's molecules. In practice, however, \emph{microscopic}
characteristics like $\Phi(r)$ are determined by calculating the corresponding
\emph{macroscopic} parameters of the fluid and fitting them to their empiric values.

Let the fluid occupy the whole space, so that the volumetric density of the
collective\ force, induced by all the molecules at a point $\mathbf{r}$, is%
\begin{equation}
\mathbf{F}(\mathbf{r},t)=\frac{\rho(\mathbf{r},t)}{m}\mathbf{\nabla}\int%
\frac{\rho(\mathbf{r}^{\prime},t)}{m}m^{2}\Phi(\mathbf{r}-\mathbf{r}^{\prime
})\,\mathrm{d}^{3}\mathbf{r}^{\prime}, \label{2.16}%
\end{equation}
where the integral in (\ref{2.16}) is to be evaluated over the whole space
(the same is implied in all further integrals with omitted limits). Expression
(\ref{2.16}) will be referred to as the Vlasov model, which is how the
collective field approach is called in plasma physics \cite{Vlasov68}.

If $\rho$ depends only on the vertical coordinate $z$ (i.e., physically, the
liquid/vapor interface is flat and horizontal), the integral in (\ref{2.16})
can be rewritten in cylindrical coordinates $\left(  r_{\bot},\beta,z^{\prime
}\right)  $ and reduced to%
\begin{equation}
F(z,t)=\rho(z,t)\frac{\partial}{\partial z}\int\rho(z^{\prime},t)\,\Psi
(z-z^{\prime})\,\mathrm{d}z^{\prime}, \label{2.17}%
\end{equation}
where%
\begin{equation}
\Psi(z)=2\pi\int_{0}^{\infty}\Phi\left(  \sqrt{r_{\bot}^{2}+z^{2}}\right)
r_{\bot}\mathrm{d}r_{\bot}. \label{2.18}%
\end{equation}
The diffuse-interface model (DIM) is based on the assumption that the spatial
scale of $\Psi(z)$ [inherited from the original intermolecular potential
$\Phi(r)$] is much shorter than that of the density field. Then, the integral
on the right-hand side of (\ref{2.17}) can be simplified by changing
$z^{\prime}\rightarrow z-z^{\prime}$, expanding $\rho(z-z^{\prime},t)$ in
powers of $z^{\prime}$, and truncating the expansion after the first three
terms. The integral involving $z^{\prime}\Psi(z^{\prime})$ vanishes (because
$\Psi(z^{\prime})=\Psi(-z^{\prime})$ due to isotropy), and one obtains
\cite{AndersonMcFaddenWheeler98,PismenPomeau00}%
\begin{equation}
F=\rho\frac{\partial}{\partial z}\left(  2a\rho+K\frac{\partial^{2}\rho
}{\partial z^{2}}\right)  , \label{2.19}%
\end{equation}
where the factor $2$ is introduced for convenience, and%
\begin{align}
a  &  =\frac{1}{2}\int\Psi(z)\,\mathrm{d}z,\label{2.20}\\
K  &  =\frac{1}{2}\int\Psi(z)\,z^{2}\mathrm{d}z \label{2.21}%
\end{align}
will be referred to as the van der Waals parameter and Korteweg parameter,
respectively. The former is the same as its namesake $a$ in the Enskog--Vlasov
fluid model (\ref{2.7})--(\ref{2.9}) -- hence, its value for a particular
fluid can be deduced by fitting the Enskog--Vlasov equation of state to its
empiric counterpart (for water, this is done in Appendix \ref{appA.1}). The
value of the Korteweg parameter $K$, in turn, can be deduced from the fluid's
capillary properties (for water, see Appendix \ref{appA.2}).

Even though $a$ and $K$ were introduced as parameters of the DIM, they can be
viewed as global characteristics of the general Vlasov potential $\Phi(r)$.
Furthermore, since they are \emph{integral} characteristics of $\Phi(r)$, they
are more important than its actual shape. In particular, Eqs. (\ref{2.20}%
)--(\ref{2.21}) suggest that the spatial scale of the vdW force is%
\begin{equation}
l_{F}=\sqrt{\frac{K}{a}}, \label{2.22}%
\end{equation}
which is one of the crucial characteristics in interfacial dynamics.

In the present paper, the following example of the Vlasov potential is used%
\begin{equation}
\Phi(r)=\left[  B\left(  1-\frac{r^{2}}{\Delta^{2}}\right)  +C\left(
1-\frac{r^{4}}{\Delta^{4}}\right)  \right]  \operatorname*{H}(\Delta-r),
\label{2.23}%
\end{equation}
where $\operatorname*{H}(\Delta-r)$ is the Heaviside step function, and $B$,
$C$, and $\Delta$ are adjustable constants. Substituting (\ref{2.23}) into
(\ref{2.18}), one obtains%
\begin{multline}
\Psi(z)=\pi\Delta^{2}\left(  1-\frac{z^{2}}{\Delta^{2}}\right)  ^{2}\\
\times\left[  \frac{B}{2}+\frac{C}{3}\left(  2+\frac{z^{2}}{\Delta^{2}%
}\right)  \right]  \operatorname*{H}(\Delta^{2}-z^{2}). \label{2.24}%
\end{multline}
To adapt $\Psi(z)$ to the fluid under consideration, one should express $B$
and $C$ through the van der Waals and Korteweg parameters, $a$ and $K$.
Substituting expression (\ref{2.24}) into (\ref{2.20})--(\ref{2.21}) and
solving for $B$ and $C$, one obtains%
\begin{align}
B  &  =\frac{105\left(  7\Delta^{2}a-45K\right)  }{16\pi\Delta^{5}%
},\label{2.25}\\
C  &  =-\frac{945\left(  \Delta^{2}a-7K\right)  }{32\pi\Delta^{5}}.
\label{2.26}%
\end{align}
With $a$ and $K$ deduced from empiric data, the undetermined parameter
$\Delta$ can be viewed as the one describing the shape of $\Phi(r)$. In
particular, $\Phi(r)$ is monotonic only if%
\begin{equation}
\sqrt{\frac{45}{7}}l_{F}\leq\Delta\leq3l_{F}. \label{2.27}%
\end{equation}
If $\Delta$ is outside this range, the vdW force (which is $\sim
\mathbf{\nabla}\Phi$) is \emph{repulsive} for some $r$ -- whereas physically,
it should be \emph{attractive}. Thus, $\Delta$ should better be chosen from
range (\ref{2.27}).

One might also argue that the vdW force should not involve a small-$r$
component: the short-range part of the intermolecular interaction is
responsible for collisions and supposed to be accounted for by the viscosity
term (in hydrodynamics) or collision integral (in kinetic theory) -- not the
Vlasov term. With this in mind, one should make $\mathbf{\nabla}\Phi$ near
$r=0$ as small as possible; in terms of expression (\ref{2.23}), this
corresponds to $B=0$, so that (\ref{2.25}) yields%
\begin{equation}
\Delta=\sqrt{\frac{45}{7}}l_{F}. \label{2.28}%
\end{equation}
Evidently, this value is included in range (\ref{2.27}) as its left-hand endpoint.

Since particular case (\ref{2.28}) satisfies all the criteria, it will be used
in the remainder of this paper.

The following comments are in order:

\begin{itemize}
\item Even if one chooses a different $\Delta$ from range (\ref{2.27}), the
fluid's macroscopic properties remain virtually the same. In particular, when
$a$ and $K$ are fixed while $\Delta$ varies through the whole range
(\ref{2.27}), the corresponding change of the surface tension $\gamma$ is
approximately $0.1\%$ (this calculation was carried out using the
Enskog--Vlasov equation of state with the parameters of water at $25^{\circ
}\mathrm{C}$; see also Appendix \ref{appA.2} for the dependence of $\gamma$ on
the equation of state, $a$, and $K$).

\item The choice of the formula for $\Phi(r)$ appears to also be unimportant:
an exponential alternative to (\ref{2.23}) was tested, and the dependence of
$\gamma$ on the temperature was found to be virtually the same.
\end{itemize}

Note that the diffuse-interface model corresponds to the limit $\Delta
\rightarrow0$ and, thus, is not included in interval (\ref{2.27}). Another
problem with the derivation of the DIM via expansion (\ref{2.19}) has been
noted in Ref. \cite{Pismen02}: for some intermolecular potentials, the
higher-order terms omitted from (\ref{2.19}) involve divergent integrals. For
potential (\ref{2.23}), this problem does not arise -- but the mere
possibility of a divergence may indicate \textquotedblleft a qualitative
difference between the solutions of the exact and truncated
equations\textquotedblright\ \cite{Pismen02}.

All this does not mean that the DIM should be discarded; it has been used for
more than a century by hundreds of researchers -- and deserves to be examined
at face value. This is what is done in the present paper, and the results
obtained are tested against those of the more general Vlasov model.

\subsection{Governing equations\label{sec2.4}}

Evaporation of common liquids at normal conditions is a slow process -- hence,
the slow-flow approximation can be used, which amounts to the following
equations:%
\begin{equation}
\frac{\partial\rho}{\partial t}+\frac{\partial\left(  \rho w\right)
}{\partial z}=0, \label{2.29}%
\end{equation}%
\begin{equation}
\frac{\partial\hat{p}}{\partial z}=\frac{\partial}{\partial z}\left(  \mu
\frac{\partial w}{\partial z}\right)  +F, \label{2.30}%
\end{equation}
where $w(z,t)$ is the vertical velocity, $\hat{p}$ is the thermal pressure,
and $\mu$ is the effective viscosity related to the shear and bulk viscosities
by%
\begin{equation}
\mu=\frac{4}{3}\mu_{s}+\mu_{b}. \label{2.31}%
\end{equation}
The non-thermal part of the pressure comes from the van der Waals force: in
the DIM, this can be shown explicitly (by substituting expression (\ref{2.19})
for $F$ into the momentum equation (\ref{2.30}) and incorporating the term
$\sim a$ into the pressure term). In the general Vlasov model, however, the
non-thermal pressure has to remain `hidden' inside the force term.

The viscosity $\mu(\rho,T)$ should be treated as a known function. Recall also
that, according to both experiment \cite{LindstromMallard97} and kinetic
theory \cite{FerzigerKaper72}, the ideal-gas limit of $\mu$ is finite, i.e.,%
\[
\mu(\rho,T)\sim\mu_{0}(T)\qquad\text{as}\qquad\rho\rightarrow0.
\]
As shown in Refs. \cite{Benilov22a,Benilov23c,Benilov23d} for the DIM,
evaporation of common liquids under normal conditions is not sensitive to the
\emph{finite-}$\rho$ range of the viscosity function $\mu(\rho,T)$; its only
important characteristic is the \emph{low}-density limit $\mu_{0}$. The same
is true for the Vlasov model (more details given below) -- thus, when
illustrating the general results, the simplest (density-independent)
approximation of the viscosity will be used, $\mu(\rho,T)=\mu_{0}(T)$. For
specific computations, one still needs the dependence of $\mu_{0}$ on $T$ --
in this paper, that of water is used (see Appendix \ref{appA.3}).

\subsection{Boundary conditions\label{sec2.5}}

Consider a flat horizontal interface separating a liquid and its vapor (the
former is located below the latter). Mathematically, this corresponds to%
\begin{equation}
\rho\rightarrow\rho^{(l)}\qquad\text{as}\qquad z\rightarrow-\infty,
\label{2.32}%
\end{equation}%
\begin{equation}
\rho\rightarrow\rho^{(v)}\qquad\text{as}\qquad z\rightarrow+\infty,
\label{2.33}%
\end{equation}
where $\rho^{(v)}$ and $\rho^{(l)}$ are the vapor and liquid densities,
respectively, and the $z$ axis is directed upwards. Note that $\rho^{(v)}$ is
a given parameter (determined by the relative humidity), whereas $\rho^{(l)}$
is to be found.

If the vapor is undersaturated ($\rho^{(v)}<\rho^{(v.sat)}$), evaporation
gives rise to a flow -- such that%
\begin{equation}
w\rightarrow0\qquad\text{as}\qquad z\rightarrow-\infty, \label{2.34}%
\end{equation}%
\begin{equation}
\frac{\partial w}{\partial z}\rightarrow0\qquad\text{as}\qquad z\rightarrow
+\infty, \label{2.35}%
\end{equation}
i.e., physically, the liquid far below the interface is at rest, and the vapor
flow far above the interface is stress-free.

Equations (\ref{2.17}), (\ref{2.29})--(\ref{2.30}) and conditions
(\ref{2.32})--(\ref{2.35}) form a boundary-value problem to be solved.

\subsection{The isobaricity condition\label{sec2.6}}

Under the DIM, one can readily show that equation (\ref{2.30}) and boundary
conditions (\ref{2.32})--(\ref{2.35}) imply that the pressure values far above
and far below the interface are equal, i.e.,%
\begin{equation}
p(\rho^{(l)},T)=p(\rho^{(v)},T). \label{2.36}%
\end{equation}
For the Vlasov model, this result holds too, but is a little harder to prove
(see Appendix \ref{appB}).

Equality (\ref{2.36}) will be referred to as the isobaricity condition; it
allows one to calculate the density $\rho^{(l)}$ of the evaporating liquid
without solving the governing equations. To do so, recall that the vapor
density $\rho^{(v)}$ is a known parameter, and treat (\ref{2.36}) as an
equation for $\rho^{(l)}$.

The isobaricity condition has another important implication: if the
temperature is sufficiently high and/or the vapor density $\rho^{(v)}$ is
sufficiently low, Eq. (\ref{2.36}) does not have a solution for $\rho^{(l)}$
-- see an illustration in Fig. \ref{fig1}. In such cases, the liquid below the
interface cannot be homogeneous; it was conjectured in Ref. \cite{Benilov23c}
that it boils, but perhaps this phenomenon should be called cavitation.

Either way, this effect will not be discussed in further detail (because it
typically occurs at a temperature much higher than \textquotedblleft
normal\textquotedblright). One should only remember that solutions describing
steady evaporation may cease to exist when the temperature exceeds a certain threshold.

\section{Evaporation of a liquid into its vapor: the solution\label{sec3}}

Assume that evaporation is steady -- hence, the liquid/vapor interface recedes
at a constant velocity equal to $-E/\rho^{(l)}$, where $E$ is the evaporation
rate and $\rho^{(l)}$, the liquid density. Thus, seek a solution of the form%
\[
\rho=\rho(z_{new}),\qquad w=w(z_{new}),
\]
where%
\[
z_{new}=z+\frac{E}{\rho^{(l)}}t.
\]
In terms of the new variable, the density equation (\ref{2.29}) becomes (the
subscript $_{new}$ omitted)%
\[
\frac{E}{\rho^{(l)}}\frac{\partial\rho}{\partial z}+\frac{\partial\left(  \rho
w\right)  }{\partial z}=0.
\]
Integrating this equation and fixing the constant via boundary conditions
(\ref{2.32}) and (\ref{2.34}), one obtains%
\begin{equation}
w=E\left(  \frac{1}{\rho}-\frac{1}{\rho^{(l)}}\right)  . \label{3.1}%
\end{equation}
Coincidently, this expression satisfies the second boundary condition for $w$,
(\ref{2.35}).

Next, rewrite the momentum equation (\ref{2.30}) in terms of $z_{new}$, omit
$_{new}$, use (\ref{3.1}) and (\ref{2.17}) to eliminate $w$ and $F$, then use
identity (\ref{2.13}) to express $\hat{p}$ through $\hat{G}$, and eventually
obtain%
\begin{multline}
\frac{\mathrm{d}}{\mathrm{d}z}\left[  \hat{G}(\rho,T)-\int\rho(z^{\prime
})\,\Psi(z-z^{\prime})\,\mathrm{d}z^{\prime}\right] \\
=-\frac{E}{\rho}\frac{\mathrm{d}}{\mathrm{d}z}\left[  \frac{\mu(\rho,T)}%
{\rho^{2}}\frac{\mathrm{d}\rho}{\mathrm{d}z}\right]  . \label{3.2}%
\end{multline}
This equation and boundary condition (\ref{2.32})--(\ref{2.33}) are invariant
with respect to an arbitrary shift $z\rightarrow z+\operatorname{const}$ --
hence, they do not fully fix the solution $\rho(z)$. An extra boundary
condition is needed -- say,%
\begin{equation}
\rho=\frac{1}{2}\left(  \rho^{(l.sat)}+\rho^{(v.sat)}\right)  \qquad
\text{at}\qquad z=0, \label{3.3}%
\end{equation}
where the saturated densities are used to `pin' solutions with different
$\rho^{(v)}$ to the same point of space associated with the equilibrium solution.

Finally, the equilibrium solution $\rho^{(sat)}(z)$ satisfies Eq. (\ref{3.2})
with $E=0$; integrating it, fixing the constant via boundary condition
(\ref{2.33}) with $\rho^{(v)}=\rho^{(v.sat)}$, and recalling equalities
(\ref{2.20}) and (\ref{2.12}), one can write the resulting equation in the
form%
\begin{multline}
\int\rho^{(sat)}(z^{\prime},t)\,\Psi(z-z^{\prime})\,\mathrm{d}z^{\prime}\\
=\hat{G}(\rho^{(sat)},T)-G(\rho^{(v.sat)},T). \label{3.4}%
\end{multline}

\subsection{The diffuse-interface model\label{sec3.1}}

The DIM equation for steady evaporation can be obtained similarly to its
Vlasov counterpart, Eq. (\ref{3.2}) -- one only needs to represent the vdW
force by the differential expression (\ref{2.19}) instead of its integral
counterpart (\ref{2.17}). Eventually, one obtains%
\begin{equation}
\frac{\mathrm{d}}{\mathrm{d}z}\left[  G(\rho,T)-K\frac{\mathrm{d}^{2}\rho
}{\mathrm{d}z^{2}}\right]  =-\frac{E}{\rho}\frac{\mathrm{d}}{\mathrm{d}%
z}\left[  \frac{\mu(\rho,T)}{\rho^{2}}\frac{\mathrm{d}\rho}{\mathrm{d}%
z}\right]  . \label{3.5}%
\end{equation}
Boundary-value problem (\ref{2.32})--(\ref{2.33}), (\ref{3.3}), (\ref{3.5})
was solved numerically using the function BVP5C of MATLAB. A typical solution
is shown, together with the equilibrium solution, in the upper panel (labeled
\textquotedblleft DIM\textquotedblright) of Fig. \ref{fig2}. The following
features should be observed:

\begin{figure}
\begin{center}\includegraphics[width=\columnwidth]{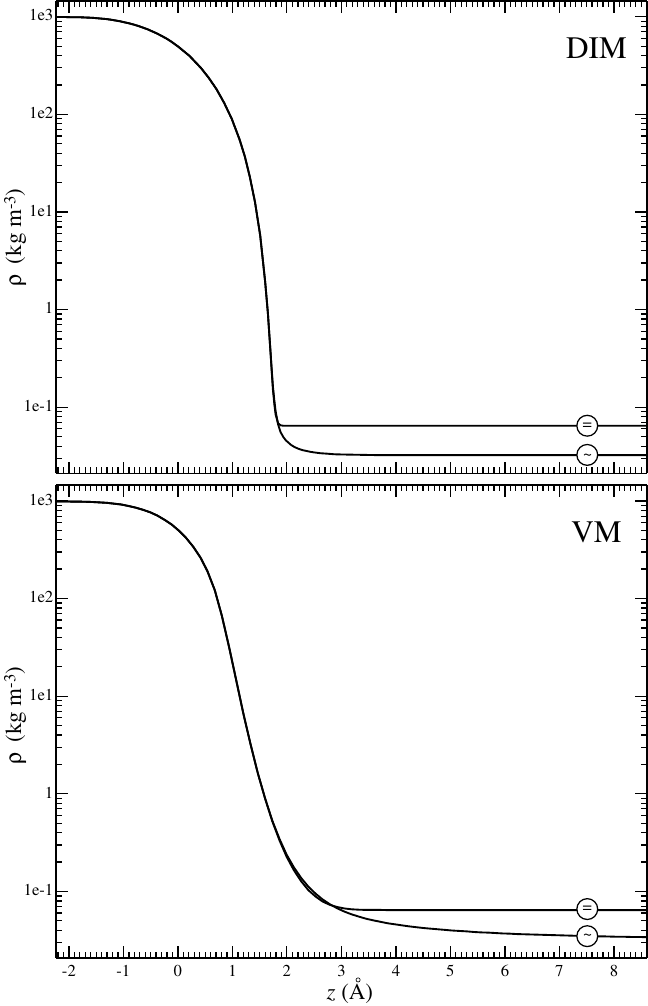}\end{center}
\caption{The equilibrium interface (marked with \textquotedblleft =\textquotedblright) and nonequilibrium interface with $H=0.5$ (marked with \textquotedblleft$\sim$\textquotedblright); in both cases, $T=50^{\circ}\mathrm{C}$. The results in panels labeled by \textquotedblleft DIM\textquotedblright\ and \textquotedblleft VM\textquotedblright\ are computed using the diffuse-interface and Vlasov models, respectively, with the parameters of water (also implied in all further figures).}
\label{fig2}
\end{figure}

\begin{itemize}
\item Within the interface, the equilibrium and non-equilibrium solutions are indistinguishable.

\item The two curves split when passing through a narrow region just outside
the interface (to be referred to as the \emph{van der Waals layer}) and remain
constant after that.
\end{itemize}

These observations help one to examine the problem asymptotically in the
low-temperature limit. Two asymptotic zones can be identified: the interface
and vdW layer. In the former, the solution is determined by the balance of the
pressure gradient and van der Waals force, whereas in the latter, viscosity
comes into play. Since the interface is, essentially, in equilibrium, it is
the vdW layer that determines the evaporation rate.

The asymptotic solution of boundary-value problem (\ref{2.32})--(\ref{2.33}),
(\ref{3.3}), (\ref{3.5}) is described in Appendix \ref{appD}, and is
summarized here in terms of the temperature $T$ and relative humidity%
\[
H=\frac{\rho^{(v)}}{\rho^{(v.sat)}}.
\]
As shown in Appendix \ref{appD}, the evaporation rate is%
\begin{equation}
E=\bar{E}_{D}(T)\,\tilde{E}_{D}(H), \label{3.6}%
\end{equation}
where%
\begin{equation}
\bar{E}_{D}(T)=\dfrac{K^{1/2}\rho^{(v.sat)5/2}\left(  RT\right)  ^{1/2}\,}%
{\mu_{0}} \label{3.7}%
\end{equation}
is of the same dimension as $E$, whereas $\tilde{E}_{D}(H)$ is nondimensional
[and determined by boundary-value problem (\ref{D.15})--(\ref{D.17})].

The exact and asymptotic solutions, both found numerically, are illustrated in
the upper panels (labeled \textquotedblleft DIM\textquotedblright) of Fig.
\ref{fig3}. The following features should be observed:

\begin{figure*}
\begin{center}\includegraphics[width=\textwidth]{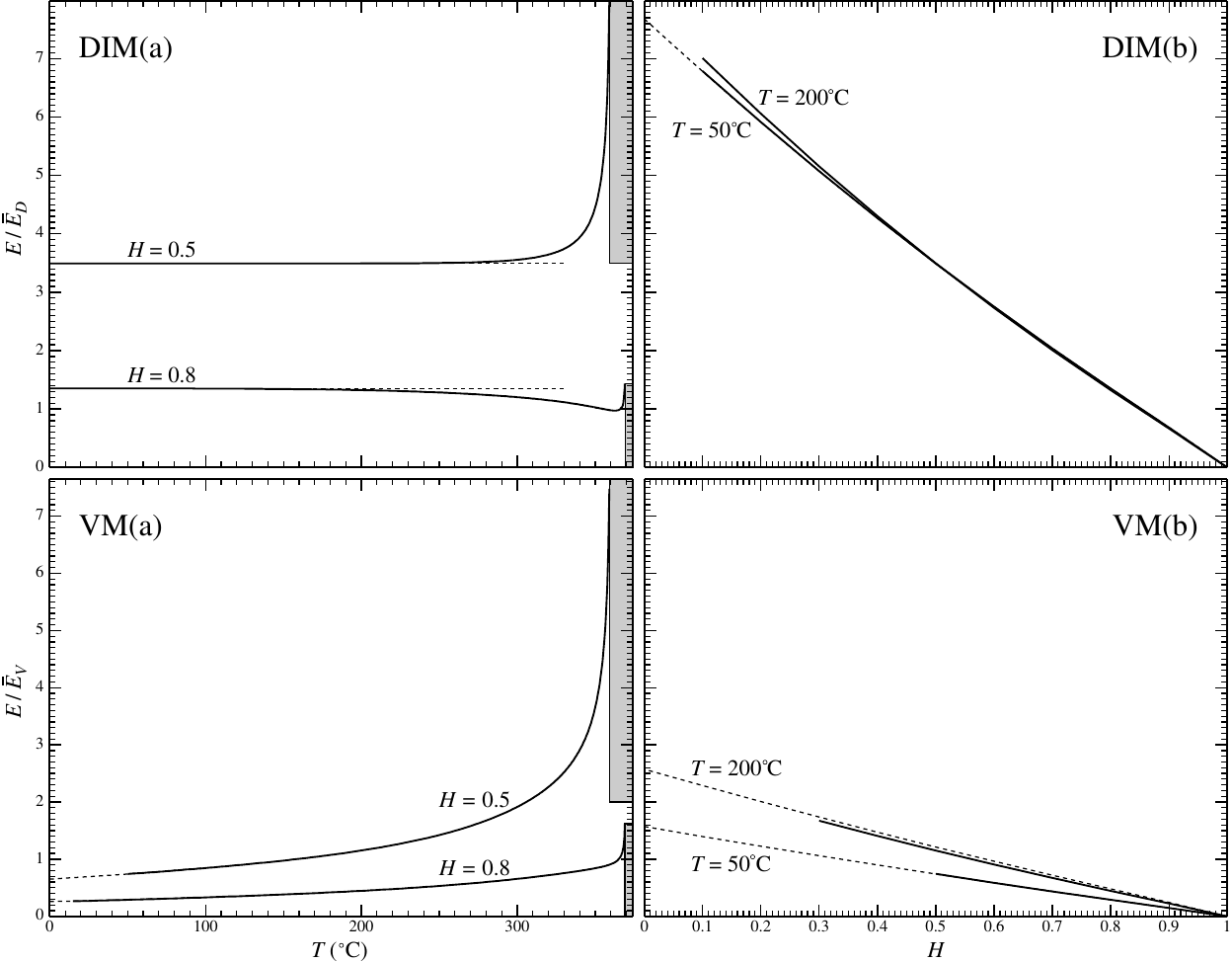}\end{center}
\caption{The evaporation rate $E$ computed via the diffuse-interface and Vlasov models, scaled by $\bar{E}_{D}$ [see (\ref{3.7})] and $\bar{E}_{V}$ [see (\ref{3.14})], respectively, are presented in the upper and lower panels, respectively. (a) $E$ vs. the temperature $T$, for two values of the relative humidity $H$; (b) $E$ vs. $H$, for two values of $T$. The solid line shows the numerical solution of the exact equations, the dotted line shows the asymptotic results obtained for the limit $\rho^{(v.sat)}/\rho^{(l.sat)} \rightarrow0$. The temperature in the two left-hand panels ranges from the triple point of water, $T\approx0^{\circ}\mathrm{C}$, to its critical point, $T\approx374^{\circ}\mathrm{C}$. The regions where the solution does not exist are shaded (their widths depend on $H$).}
\label{fig3}
\end{figure*}

\begin{itemize}
\item Panel DIM(a) shows that the asymptotic solution becomes inapplicable
near the point where the exact solution ceases to exist. This comes as no
surprise, as this temperature is fairly close to the critical point -- hence,
$\rho^{(v.sat)}/\rho^{(l.sat)}$ is not small there.

\item Panel DIM(b) shows that the exact curves for different $T$ collapse, or
nearly collapse, onto the same asymptotic curve. This is a result of the
`separation' of $T$ and $H$ in expression (\ref{3.6}).

\item The solid curves in panel DIM(b) are not extended to small $H$ because
the exact solution is difficult to compute in this region. The difficulty is
caused by the smallness of the vapor density, so that the last term in Eq.
(\ref{3.5}) becomes nearly singular.

\item The asymptotic curve in panel DIM(b) is close to a straight line
(although it is not one exactly).
\end{itemize}

As shown in Appendix \ref{appD}, the spatial scale of the vdW layer, $l_{L}$,
and that of the vdW force, $l_{F}$ [given by (\ref{2.22})], are such that%
\begin{equation}
\frac{l_{L}}{l_{F}}=T_{nd}^{-1/2}\left(  \frac{\rho^{(v.sat)}}{\rho^{(l.sat)}%
}\right)  ^{1/2}, \label{3.8}%
\end{equation}
where $T_{nd}$ is the nondimensional temperature defined by (\ref{2.15}).
Recalling that, for common liquids under normal conditions, $T_{nd}$ is small,
whereas $\rho^{(v.sat)}/\rho^{(l.sat)}$ is \emph{exponentially} small, one
concludes that $l_{L}\ll l_{F}$. This is the opposite of what the DIM is
applicable to.

This observation motivated the author of the present paper to re-examine the
problem using the Vlasov model (whose applicability does not require that the
flow's spatial scale be large).

\subsection{The Vlasov model\label{sec3.2}}

The Vlasov equation (\ref{3.2}) is much harder to solve numerically than its
DIM counterpart (\ref{3.5}): the former is of an \emph{integro}-differential
kind, for which there are no ready-made tools in MATLAB or similar packages.
The only numerical algorithm the author of this paper has come up with (see
Appendix \ref{appC}) does not perform well for small $H$ and/or $T$, and often
requires manual fine-tuning of the computational parameters.

A typical solution of Eq. (\ref{3.2}) subject to conditions (\ref{2.32}%
)--(\ref{2.33}) and (\ref{3.3}) is shown in the lower panel (labeled
\textquotedblleft VM\textquotedblright) of Fig. \ref{fig2}. The interfacial
region is evidently close to equilibrium, whereas evaporation occurs in the
vdW layer -- just like it does under the DIM. The two models are still
different, however: the width of the vdW layer in the Vlasov model is
comparable to that of the interface, not smaller. This is visible in the lower
panel of Fig. \ref{fig2}, but can also be deduced from Eq. (\ref{3.2}) directly.

To do so, observe that the main contribution to the integral term in
(\ref{3.2}) comes from the interfacial region (where $\rho$ is large), whereas
the contribution of the vdW layer (small $\rho$) is negligible. Recalling also
that the former region is in equilibrium, one can approximate Eq. (\ref{3.2})
by%
\begin{multline}
\frac{\mathrm{d}}{\mathrm{d}z}\left[  \hat{G}(\rho,T)-\int\rho^{(sat)}%
(z^{\prime})\,\Psi(z-z^{\prime})\,\mathrm{d}z^{\prime}\right] \\
=-\frac{E}{\rho}\frac{\mathrm{d}}{\mathrm{d}z}\left[  \frac{\mu(\rho,T)}%
{\rho^{2}}\frac{\mathrm{d}\rho}{\mathrm{d}z}\right]  , \label{3.9}%
\end{multline}
where $\rho^{(sat)}(z)$ is the equilibrium solution.

This equation can be simplified in two different ways. First, recall that
$\rho^{(sat)}(z)$ satisfies Eq. (\ref{3.4}) -- hence, the integral term in Eq.
(\ref{3.9}) can be replaced with $G(\rho^{(v.sat)},T)-\hat{G}(\rho^{(sat)}%
,T)$. Second, assume that $z$ is inside the vdW layer -- hence, $\rho(z)$ is
small. This allows one to replace $\mu(\rho,T)$ with its low-density limit
$\mu_{0}(T)$, and $\hat{G}(\rho,T)$ and $\hat{G}(\rho^{(sat)},T)$, with\ their
low-density asymptotics (\ref{2.14}). Eventually, the original \emph{integro}%
-differential equation reduces to a \emph{differential} one,%
\begin{multline}
RT\left(  \frac{1}{\rho}\frac{\mathrm{d}\rho}{\mathrm{d}z}-\frac{1}%
{\rho^{(sat)}}\frac{\mathrm{d}\rho^{(sat)}}{\mathrm{d}z}\right) \\
=-\frac{E\mu_{0}(T)}{\rho}\frac{\mathrm{d}}{\mathrm{d}z}\left(  \frac{1}%
{\rho^{2}}\frac{\mathrm{d}\rho}{\mathrm{d}z}\right)  . \label{3.10}%
\end{multline}
Clearly, the solution $\rho(z)$ has the same spatial scale as $\rho
^{(sat)}(z)$ -- simply because Eq. (\ref{3.10}) does not involve other spatial scales.

Equation (\ref{3.10}) is to be solved with boundary condition (\ref{2.33}),
rewritten in terms of the relative humidity and saturated vapor density,%
\begin{equation}
\rho\rightarrow H\rho^{(v.sat)}\qquad\text{as}\qquad z\rightarrow\infty.
\label{3.11}%
\end{equation}
The boundary condition at the other end is unclear, however -- and this is not
a technical glitch, but a fundamental issue. It results from the fact that, in
the problem under consideration, the neighboring asymptotic zones are on the
same spatial scale and, thus, cannot be matched via the van Dyke rule or a
similar method.

This difficulty can probably be resolved by changing the variables $\left(
z,\rho\right)  \rightarrow\left(  \rho,q=\mathrm{d}\rho/\mathrm{d}z\right)  $,
in which case the interface would correspond to $\rho\sim\rho^{(l.sat)}$ and
the vdW layer, to $\rho\sim\rho^{(v.sat)}$. This is, essentially, how matching
was handled under the DIM (see Appendix \ref{appD}) -- but, in the Vlasov
model, the new variables complicate the integral representing the vdW force.

In the end, the following workaround was used. A particular point $z_{m}$ was
picked, such that%
\begin{equation}
\rho^{(v.sat)}\ll\rho^{(sat)}(z_{m})\ll\rho^{(l.sat)}, \label{3.12}%
\end{equation}
and the vdW layer solutions was `patched' at $z=z_{m}$ to the interfacial
(equilibrium) solution,%
\begin{equation}
\rho=\rho^{(sat)},\qquad\frac{\mathrm{d}\rho}{\mathrm{d}z}=\frac
{\mathrm{d}\rho^{(sat)}}{\mathrm{d}z}\qquad\text{at}\qquad z=z_{m}.
\label{3.13}%
\end{equation}
With $\rho^{(sat)}(z)$ known (pre-computed), boundary-value problem
(\ref{3.10})--(\ref{3.11}), (\ref{3.13}) fully determines the asymptotic
solution $\rho(z)$. Note that the `patching' has been previously used in other
problems, and the results have been shown to be asymptotically equivalent to
those obtained via matching \cite{BenilovM03}.

Boundary-value problem (\ref{3.10})--(\ref{3.11}), (\ref{3.13}) was solved
numerically using the function BVP5C of MATLAB. The patching point $z_{m}$ was
chosen such that%
\[
\rho^{(sat)}(z_{m})=\sqrt{\rho^{(v.sat)}\rho^{(l.sat)}},
\]
which automatically satisfies restrictions (\ref{3.12}). The computed
evaporation rate $E$ can be written in a nondimensional form, relative to%
\begin{equation}
\bar{E}_{V}=\frac{K^{1/2}\rho^{(v.sat)2}RT}{a^{1/2}\mu_{0}}, \label{3.14}%
\end{equation}
in which case $E/\bar{E}_{V}$ happens to be order one (provided $T$ is not too
close to the point where the solution ceases to exist).

Typical numerical results are illustrated in the lower panels (labeled
\textquotedblleft VM\textquotedblright) of Fig. \ref{fig3}. The following
features should be observed:

\begin{itemize}
\item The solid curves in panel VM(b) are not extended to small $H$ because
the exact boundary-value problem computes much worse than its asymptotic counterpart.

\item Panel VM(a) of Fig. \ref{fig3} illustrates that, for mid-range $T$ and
$H$, the asymptotic approach works well (the asymptotic curves in this figure
are actually drawn for $T\leq100^{\circ}\mathrm{C}$, but they are
indistinguishable from the exact solution).

\item Panel VM(b) illustrates that the asymptotic and exact solutions start to
diverge near $T=200^{\circ}\mathrm{C}$.

\item Observe that the curves corresponding to different $T$ in panel VM(b) do
\emph{not} collapse onto a single curve [unlike those computed via the DIM and
illustrated in panel DIM(b)]. In principle, this could be a result of choosing
the wrong scale $\bar{E}_{V}$ -- and so some other were tested, but none
worked. One might conclude that, for the Vlasov model, $E$ depends on $T$ and
$H$ in a non-separable way.

\item The curves in panel VM(b) of Fig. \ref{fig3} look like straight lines
(but are not ones exactly).
\end{itemize}

Thus, the DIM and VM both predict an almost linear dependence of $E$ on $H$,
which allows one to compare them to the Hertz--Knudsen Law (HKL).

\subsection{DIM and VM vs. HKL\label{sec3.3}}

Rewrite the Hertz--Knudsen Law, given by Eq. (\ref{1.2}), in terms of the
relative humidity%
\[
E=\theta\,\sqrt{\frac{RT}{2\pi}}\rho^{(v.sat)}\left(  1-H\right)  .
\]
To compare this expression to the corresponding results of the DIM and VM, the
two latter models should be represented in a similar fashion -- say,%
\begin{equation}
E(T,H)=\bar{E}(T)\,\left(  1-H\right)  . \label{3.15}%
\end{equation}
Once this representation is in place, one can \emph{calculate} the
evaporation/condensation probability,%
\begin{equation}
\theta=\frac{\bar{E}}{\rho^{(v.sat)}}\sqrt{\frac{2\pi}{RT}}, \label{3.16}%
\end{equation}
as opposed to just \emph{inserting it into the HKL and treating as an
adjustable parameter}.

One way to obtain a formula for $E$ of form (\ref{3.15}) consists in assuming
that $H\rightarrow1$ (the vapor is almost saturated) and expanding the DIM or
VM solutions in powers of $\left(  1-H\right)  $. In this expansion, $\bar
{E}(T)$ is the coefficient of the first term. Alternatively, one can determine
$\bar{E}(T)$ by curve fitting, but such an approach is less `clean' than the asymptotics.

In application to the diffuse-interface model, the expansion in $\left(
1-H\right)  $ can be found in Ref. \cite{Benilov22a}, and for the Vlasov
model, in Appendix \ref{appE} of the present paper. Under an additional
assumption that $\rho^{(v.sat)}/\rho^{(l.sat)}\ll1$, one obtains%
\begin{equation}
\bar{E}=\frac{KRT}{a\mu_{0}}\left[  \int\frac{1}{\rho^{(sat)4}}\left(
\frac{\mathrm{d}\rho^{(sat)}}{\mathrm{d}z}\right)  ^{2}\mathrm{d}z\right]
^{-1}. \label{3.17}%
\end{equation}
This expression applies to both DIM and VM, but the resulting $\bar{E}$
depends on which model is used to calculate the profile of the equilibrium
interface $\rho^{(sat)}(z)$.

\begin{figure}
\begin{center}\includegraphics[width=\columnwidth]{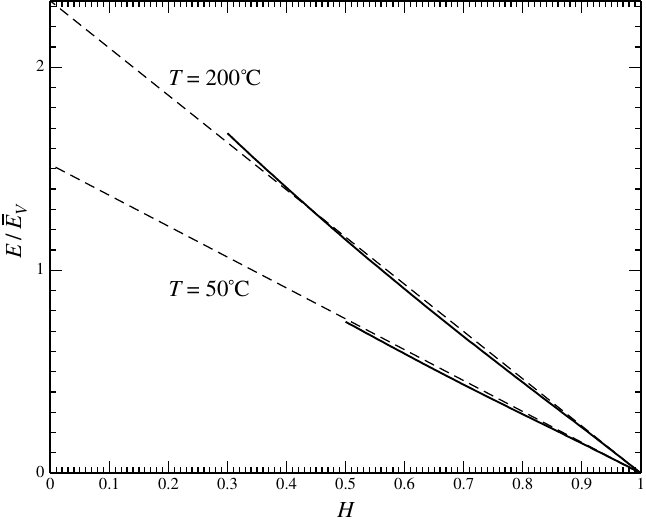}\end{center}
\caption{The evaporation rate $E$ computed via the Vlasov model and scaled by $\bar{E}_{V}$ [given by (\ref{3.14})] vs. $H$, for different values of $T$. The solid line shows the numerical solution of the exact equations (the same as in panel VM(b) of Fig. \ref{fig3}), the dashed line shows the asymptotic results obtained for the limit $H\rightarrow1$.}
\label{fig4}
\end{figure}

In application to the Vlasov model, formulae (\ref{3.15}) and (\ref{3.17}) are
illustrated in Fig. \ref{fig4}. One can see that $\bar{E}$ approximates the
slope of the exact curve reasonably accurately.

The dependence of $\theta$ on $T$, calculated via formulae (\ref{3.16}%
)--(\ref{3.17}), is illustrated for water in Fig. \ref{fig5}. The following
features should be observed:

\begin{figure}
\begin{center}\includegraphics[width=\columnwidth]{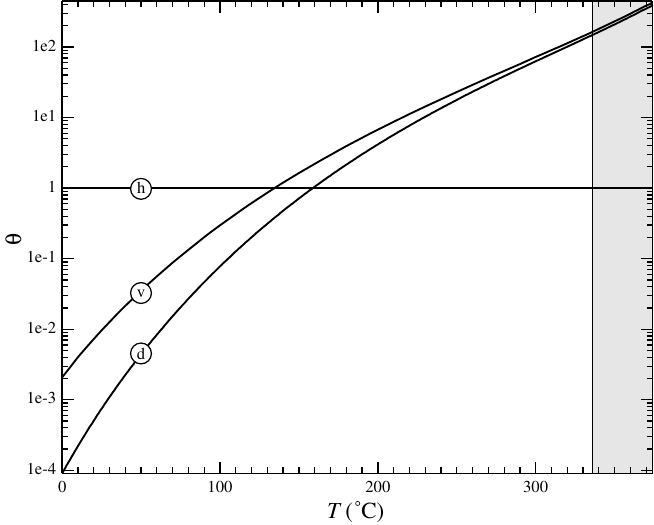}\end{center}
\caption{The evaporation/condensation probability $\theta$ vs. $T$. Curves (d) and (v) correspond to the DIM and VM, respectively. The straight line marked (h) corresponds to $\theta=1$, as in the original assumption of Hertz and Knudsen \cite{Hertz82,Knudsen15}.}
\label{fig5}
\end{figure}

\begin{itemize}
\item For $T<100^{\circ}\mathrm{C}$, the DIM and VM both predict that $\theta$
is small (in contrast to the classical formulation of the HKL, where
$\theta=1$).

\item For $T>250^{\circ}\mathrm{C}$, the DIM and VM both predict that $\theta$
is large. This conclusion, however, can be trusted only qualitatively, as
$\rho^{(v.sat)}/\rho^{(l.sat)}$ is not necessarily small in this range (making
the approximation of viscosity employed for computation of curves (d) and (v)
invalid).\newline\hspace*{0.5cm}Note that, even though the coefficient
$\theta$ in the HKL was initially interpreted as the probability of a molecule
to evaporate or condensate, more recent theoretical models (e.g.,
\cite{Schrage53}) argue that, due to other effects, $\theta$ can exceed unity.
Thus, the present results are not surprising just because $\theta>1$, but
because $\theta\gg1$.

\item For the most of the temperature range, the DIM noticeably underpredicts
$\theta$ by comparison with the (more accurate) VM. This is a result of the
former's failure in the van der Waals layer.

\item It is worth mentioning that the slow-flow approximation [used to reduce
the exact momentum equation to Eq. (\ref{2.30})] does not impose any
restrictions on the results. Indeed, let the Reynolds number be%
\begin{equation}
Re=\frac{\rho^{(v.sat)}\bar{v}l_{F}}{\mu},\label{3.18}%
\end{equation}
where spatial scale of the interface $l_{F}$ is given by (\ref{2.22}) and the
velocity scale can be expressed through the evaporation rate $\bar{E}_{V}$ of
the Vlasov model [given by (\ref{3.14})],%
\begin{equation}
\bar{v}=\frac{\bar{E}_{V}}{\rho^{(v.sat)}}.\label{3.19}%
\end{equation}
Estimating expressions (\ref{3.18})--(\ref{3.19}) for water, one can show that
$Re$ is consistently small for all $T$ between the triple and critical points
(reaching the maximum at the latter, where $Re\approx0.022$).
\end{itemize}

\section{Evaporation of a liquid into air: the formulation\label{sec4}}

\subsection{Thermodynamics\label{sec4.1}}

Following Refs. \cite{Benilov23a,Benilov23d}, air will be treated as a single
fluid with its parameters equal to the 79/21 weighted averages of those of
nitrogen and oxygen. Thus, the problem will be formulated for a two-component
fluid, representing either a liquid with air dissolved in it, or a mixture of
vapor and air.

The thermodynamic state of a multicomponent fluid can be characterized by the
temperature $T$ and partial densities $\rho_{i}$ ($i=1$ represents the liquid
or its vapor, and $i=2$ represents the air). Introducing the specific internal
energy $e(\rho_{1},\rho_{2},T)$ and entropy $s(\rho_{1},\rho_{2},T)$
[satisfying the Gibbs relation (\ref{2.1})], one can define the full pressure
and chemical potentials by%
\begin{align}
p  &  =\rho\sum_{i}\rho_{i}\left(  \frac{\partial e}{\partial\rho_{i}}%
-T\frac{\partial s}{\partial\rho_{i}}\right)  ,\label{4.1}\\
G_{i}  &  =\frac{\partial\left(  \rho e\right)  }{\partial\rho_{i}}%
-T\frac{\partial\left(  \rho s\right)  }{\partial\rho_{i}}, \label{4.2}%
\end{align}
where%
\begin{equation}
\rho=\rho_{1}+\rho_{2} \label{4.3}%
\end{equation}
is the full density. It can be readily deduced from Eqs. (\ref{4.1}%
)--(\ref{4.3}) that%
\begin{equation}
\frac{\partial p}{\partial\rho_{j}}=\sum_{i}\rho_{i}\frac{\partial G_{i}%
}{\partial\rho_{j}}. \label{4.4}%
\end{equation}
This equality is the multicomponent analogue of the pure-fluid identity
(\ref{2.4}), and Eqs. (\ref{4.1})--(\ref{4.2}) are those of Eqs.
(\ref{2.2})--(\ref{2.3}).

The multicomponent Maxwell construction, in turn, is%
\begin{equation}
p(\rho_{1}^{(a.sat)},\rho_{2}^{(a.sat)},T)=p(\rho_{1}^{(l.sat)},\rho
_{2}^{(l.sat)},T), \label{4.5}%
\end{equation}%
\begin{equation}
G_{1}(\rho_{1}^{(a.sat)},\rho_{2}^{(a.sat)},T)=G_{1}(\rho_{1}^{(l.sat)}%
,\rho_{2}^{(l.sat)},T), \label{4.6}%
\end{equation}%
\begin{equation}
G_{2}(\rho_{1}^{(a.sat)},\rho_{2}^{(a.sat)},T)=G_{2}(\rho_{1}^{(l.sat)}%
,\rho_{2}^{(l.sat)},T), \label{4.7}%
\end{equation}
where $^{a.sat}$ stands for saturated air. To fix the \emph{four} unknowns --
$\rho_{1}^{(a.sat)}$, $\rho_{2}^{(a.sat)}$,$~\rho_{1}^{(l.sat)}$, and
$\rho_{2}^{(l.sat)}$ -- the above \emph{three} equations should be
complimented with the requirement that the saturated air pressure be equal to
its atmospheric value,%
\begin{equation}
p(\rho_{1}^{(a.sat)},\rho_{2}^{(a.sat)},T)=1\,\mathrm{atm}. \label{4.8}%
\end{equation}
Next, the multicomponent version of the Enskog--Vlasov fluid model consists in%
\begin{align}
e  &  =\frac{T}{\rho}\sum_{i}c_{i}\rho_{i}-\frac{1}{\rho}\sum_{i,j}a_{ij}%
\rho_{i}\rho_{j},\label{4.9}\\
s  &  =\frac{\ln T}{\rho}\sum_{i}c_{i}\rho_{i}-\frac{1}{\rho}\sum_{i}R_{i}%
\rho_{i}\ln\rho_{i}-\Theta(\rho_{1},\rho_{2}), \label{4.10}%
\end{align}
where $c_{i}$ is the specific heat capacity of the $i$-th components, $R_{i}$
is its specific gas constant, $a_{ij}$ is the van der Waals coefficient
describing the interaction of the $i$-th and $j$-th components, and
$\Theta(\rho_{1},\rho_{2})$ is the non-ideal part of the entropy. The values
of $c_{i}$ and $R_{i}$ can be found in thermodynamics handbooks, and $a_{ij}$
and $\Theta(\rho_{1},\rho_{2})$ should be fitted to the thermodynamic
properties of the multicomponent fluid under consideration (See Appendix
\ref{appA.1}).

The thermal pressure and thermal chemical potentials are related to the full
ones by%
\begin{equation}
\hat{p}=p+\sum_{i,j}a_{ij}\rho_{i}\rho_{j},\qquad\hat{G}_{i}=G_{i}+2\sum
_{j}a_{ij}\rho_{j}. \label{4.11}%
\end{equation}
Using these expressions and identity (\ref{4.4}), one can verify that%
\begin{equation}
\frac{\partial\hat{p}}{\partial\rho_{j}}=\sum_{i}\rho_{i}\frac{\partial\hat
{G}_{i}}{\partial\rho_{j}}. \label{4.12}%
\end{equation}
Finally, the low-density asymptotics of the full and thermal chemical
potentials are both given by the ideal gas formula,%
\begin{equation}
\hat{G}_{i}\sim R_{i}T\ln\rho_{i}\qquad\text{as}\qquad\rho_{1},\rho
_{2}\rightarrow0. \label{4.13}%
\end{equation}
This expression applies only if both $\rho_{1}$ and $\rho_{2}$ are small, so
both fluids can be treated as ideal gases.

\subsection{Governing equations\label{sec4.2}}

Let $\Psi_{ij}(z)$ be the one-dimensional potential of the vdW force exerted
by component $i$ on component $j$ and vice versa ($\Psi_{ij}=\Psi_{ji}$). The
van der Waals and Korteweg parameters are now matrices, given by%
\begin{equation}
a_{ij}=\frac{1}{2}\int\Psi_{ij}(z)\,\mathrm{d}z, \label{4.14}%
\end{equation}%
\[
K_{ij}=\frac{1}{2}\int\Psi_{ij}(z)\,z^{2}\mathrm{d}z.
\]
In what follows, the general results will be illustrated using $\Psi_{ij}$
described by formulae (\ref{2.24})--(\ref{2.26}) and (\ref{2.28}), with $a$
and $K$ changed to $a_{ij}$ and $K_{ij}$ (their values for the water--air
combination are described in Appendices \ref{appA.1}--\ref{appA.2}).

The equations governing slow isothermal dynamics of a binary mixture are%
\begin{equation}
\frac{\partial\rho_{1}}{\partial t}+\frac{\partial}{\partial z}\left(
\rho_{1}w+J\right)  =0, \label{4.15}%
\end{equation}%
\begin{equation}
\frac{\partial\left(  \rho_{1}+\rho_{2}\right)  }{\partial t}+\frac
{\partial\left[  \left(  \rho_{1}+\rho_{2}\right)  w\right]  }{\partial z}=0,
\label{4.16}%
\end{equation}%
\begin{equation}
\frac{\partial\hat{p}}{\partial z}=\frac{\partial}{\partial z}\left(  \mu
\frac{\partial w}{\partial z}\right)  +\sum_{i}\rho_{i}F_{i}, \label{4.17}%
\end{equation}
where $w$ is the velocity of the mixture as a whole,%
\begin{equation}
F_{i}(z,t)=\frac{\partial}{\partial z}\sum_{j}\int\rho_{j}(z^{\prime}%
,t)\,\Psi_{ij}(z-z^{\prime})\,\mathrm{d}z^{\prime} \label{4.18}%
\end{equation}
is the vdW force affecting the $i$-th component,%
\begin{equation}
J=-D\left(  \frac{\partial\hat{G}_{1}}{\partial z}-F_{1}-\frac{\partial\hat
{G}_{2}}{\partial z}+F_{2}\right)  \label{4.19}%
\end{equation}
is the diffusive flux, and the diffusion coefficient $D$ is a known function
of $\rho_{1}$, $\rho_{2}$, and $T$. Observe that $J$ is expressed in terms of
the gradients of the chemical potentials, not densities: these two
representations are mathematically equivalent, but the former is more
convenient in the problem at hand (as well as some others, e.g., Refs.
\cite{GiovangigliMatuszewski13,LiuAmbergDoquang16}).

To establish the correspondence between $D$ and the standard diffusivity
$\mathcal{D}$ which appears in Fick's Law, one needs to adapt the
diffusive-flux expression (\ref{4.19}) to the ideal-gas limit: set $F_{i}=0$
(no vdW force) and replace $\hat{G}_{i}$ with its small-density asymptotics
(\ref{4.13}). One should also assume that the density of air exceeds that of
vapor ($\rho_{2}\gg\rho_{1}$) but their gradients are comparable
($\partial\rho_{2}/\partial z\sim\partial\rho_{1}/\partial z$). Eventually,
one obtains, to leading order,%
\[
J=-\underset{\mathcal{D}}{\underbrace{D\frac{TR_{1}}{\rho_{1}}}}\frac
{\partial\rho_{1}}{\partial z}.
\]
Comparing this equality to the standard formulation of Fick's Law, one can see
that%
\begin{equation}
D=\frac{\rho_{1}}{R_{1}T}\mathcal{D}. \label{4.20}%
\end{equation}
This formula applies only in the low-density limit, which is sufficient for
the calculations below.

Using identity (\ref{4.12}), one can rewrite the momentum equation
(\ref{4.17}) in the form%
\begin{equation}
\sum_{i}\rho_{i}\left(  \frac{\partial\hat{G}_{i}}{\partial z}-F_{i}\right)
=\frac{\partial}{\partial z}\left(  \mu\frac{\partial w}{\partial z}\right)  .
\label{4.21}%
\end{equation}

\subsection{Boundary conditions\label{sec4.3}}

The solution of the governing equations should satisfy%
\begin{equation}
\rho_{i}\rightarrow\rho_{i}^{(l)}\qquad\text{as}\qquad z\rightarrow-\infty,
\label{4.22}%
\end{equation}%
\begin{equation}
\rho_{i}\rightarrow\rho_{i}^{(a)}\qquad\text{as}\qquad z\rightarrow+\infty,
\label{4.23}%
\end{equation}
where $\rho_{i}^{(l)}$ and $\rho_{i}^{(a)}$ are the partial densities of the
$i$-th component in the liquid and air, respectively. $\rho_{1}^{(a)}$ and
$\rho_{2}^{(a)}$ should be treated as given parameters (reflecting the
humidity and air density), whereas $\rho_{1}^{(l)}$ and $\rho_{2}^{(l)}$ are
to be calculated together with the full solution. The isobaricity condition
(which holds for the multicomponent equations as well) is not sufficient to
fix them both.

Boundary conditions (\ref{4.22})--(\ref{4.23}) should be complimented with the
zero-velocity requirement (\ref{2.34}) for the liquid far below the interface,
and the zero-viscous-stress requirement (\ref{2.35}) for the air far above the interface.

\subsection{Equilibrium interfaces\label{sec4.4}}

\begin{figure}
\begin{center}\includegraphics[width=\columnwidth]{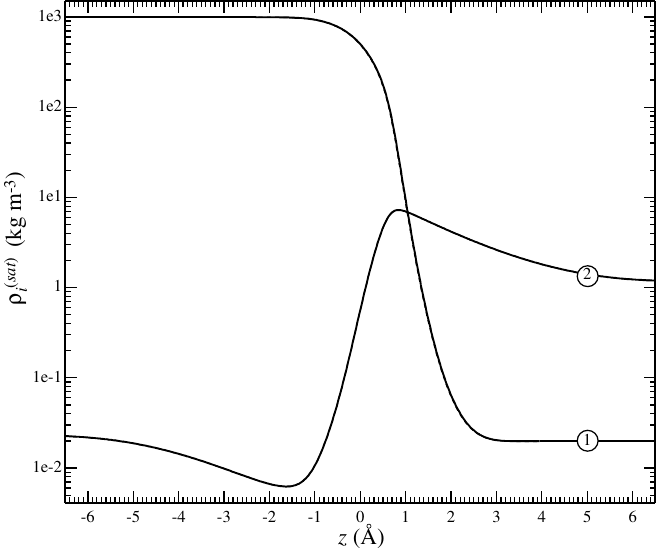}\end{center}
\caption{The equilibrium water/air interface for $T=25^{\circ }\mathrm{C}$. Curves (w) and (a) show the density of water and air, respectively. The exact and asymptotic solutions are shown in solid and dotted line, respectively, but they are virtually indistinguishable.}
\label{fig6}
\end{figure}

At equilibrium, there should be no flow ($w=0$) and no diffusive flux ($J=0$).
Substituting expression (\ref{4.18}) for $F$ into the momentum equation
(\ref{4.21}), one can integrate the latter, fix the constant of integration
via boundary condition (\ref{4.23}) with $\rho_{i}^{(a)}=\rho_{i}^{(a.sat)}$,
and recall Eqs. (\ref{4.11}) and (\ref{4.14}), to obtain%
\begin{multline}
\hat{G}_{i}(\rho_{1}^{(sat)},\rho_{2}^{(sat)},T)-\sum_{j}\int\rho_{j}%
^{(sat)}(z^{\prime})\,\Psi_{ij}(z-z^{\prime})\,\mathrm{d}z^{\prime}\\
=G_{i}(\rho_{1}^{(a.sat)},\rho_{2}^{(a.sat)},T). \label{4.24}%
\end{multline}
This equation and boundary conditions (\ref{4.22})--(\ref{4.23}) with
$\rho_{i}^{(l)}=\rho_{i}^{(l.sat)}$ and $\rho_{i}^{(a)}=\rho_{i}^{(a.sat)}$
were solved numerically using the same algorithm as that for pure fluids. A
typical solution is shown in Fig. \ref{fig6}.

The most interesting feature of liquid/air interfaces is the local maximum of
the air density. It also arises under the DIM -- see Refs.
\cite{Benilov23a,Benilov23d} where it was argued that it emerges because the
vdW force exerted by the liquid pulls extra air towards the interface. Observe
also that the interfacial width corresponding to the solution depicted in Fig.
\ref{fig6} is approximately $3\div5\mathrm{\mathring{A}}$, which agrees with
the computations/measurements reported in Refs.
\cite{LiuHarderBerne05,VerdeBolhuisCampen12,PezzottiGalimbertiGaigeot17,PezzottiServaGaigeot18,DodiaOhtoImotoNagata19}%
.

\cite{LiuHarderBerne05,VerdeBolhuisCampen12,PezzottiGalimbertiGaigeot17,PezzottiServaGaigeot18,DodiaOhtoImotoNagata19}%
.

Since the air density is much smaller than the liquid density, Eqs.
(\ref{4.24}) can be simplified asymptotically. Firstly, one can neglect the
vdW forces exerted by the air on the liquid, its vapor, and itself. Secondly,
one can neglect $\rho_{2}$ in the expression for $G_{1}$ and $\hat{G}_{1}$
(but not in $G_{2}$ and $\hat{G}_{2}$ which include $\ln\rho_{2}$). As a
result, (\ref{4.24})$_{i=1}$ reduces to the equation describing the
equilibrium liquid/vapor interface in a \emph{pure} fluid (no air involved),
and (\ref{4.24})$_{i=2}$ becomes%
\begin{multline}
\hat{G}_{2}(\rho_{1}^{(sat)},\rho_{2}^{(sat)},T)-\int\rho_{1}^{(sat)}%
(z^{\prime})\,\Psi_{12}(z-z^{\prime})\,\mathrm{d}z^{\prime}\\
=G_{2}(\rho_{1}^{(a.sat)},\rho_{2}^{(a.sat)},T). \label{4.25}%
\end{multline}
With $\rho_{1}^{(sat)}(z)$ computed from the pure-fluid problem, one can use
this equation to find $\rho_{2}^{(sat)}$. Most importantly, Eq. (\ref{4.25})
is \emph{algebraic} -- hence, a lot easier to solve numerically than the
original \emph{integro-differential} equation (\ref{4.24})$_{i=2}$.

For the parameters of water and air at normal conditions, the asymptotic and
exact solutions are virtually indistinguishable (as illustrated in Fig.
\ref{fig6}).

\subsection{Is diffusion important?\label{sec4.5}}

As shown in Ref. \cite{Benilov23a}, the importance of diffusion is
characterized by the following nondimensional parameter:%
\begin{equation}
\delta=l^{2}\frac{\bar{\rho}^{2}}{\bar{\mu}\bar{D}}, \label{4.26}%
\end{equation}
where $l$ is the characteristic interfacial width, $\bar{\rho}$ is the
characteristic density scale, $\bar{\mu}$ is the viscosity scale, etc.

Recall that evaporation of a liquid into its vapor was driven by the van der
Waals layer \emph{located just outside the interface}. The same should be
expected for evaporation of liquids into air -- hence, one needs to estimate
$\delta$ specifically for the vdW layer.

To do so, one should use the density and viscosity of air: letting, say,
$T=25^{\circ}\mathrm{C}$, one can use Refs.
\cite{TheEngineeringToolbox-AirDensity,ShangWuWangYangYeHuTaoHe19} to obtain%
\[
\bar{\rho}=1.184\,\mathrm{kg\,m}^{-3},
\]%
\[
\bar{\mu}=\left[  \frac{4}{3}\left(  1.840\times10^{-5}\right)  +\left(
1.75\times10^{-5}\right)  \right]  \mathrm{kg\,m}^{-1}\mathrm{s}^{-1}%
\]
(the decimal numbers in the latter formula represent the air's shear and bulk
viscosities). For common fluids at normal conditions, the interfacial width is
comparable to the scale $l_{F}$ of the vdW force. The latter is given by
(\ref{2.22}) -- hence,%
\[
l=\sqrt{\frac{K_{11}}{a_{11}}},
\]
where with $K_{11}$ and $a_{11}$ are the Korteweg and vdW parameters of the
liquid. For water (see Appendix \ref{appA}), one obtains $l\approx
1\mathrm{\mathring{A}}$, which qualitatively agrees with the spatial scale one
can observe in Fig. \ref{fig6}.

The diffusion coefficient $\bar{D}$, in turn, will be estimated via its
low-density asymptotics (\ref{4.20}) with%
\[
\rho_{1}=0.023075\,\mathrm{kg\,m}^{-3}%
\]
(which is the density of saturated water vapor at $25^{\circ}\mathrm{C}$
according to Ref. \cite{LindstromMallard97}) and%
\[
\mathcal{D}=2.49\times10^{-5}\mathrm{m}^{2}\mathrm{s}^{-1}%
\]
(which is the diffusivity of water vapor in air at $25^{\circ}\mathrm{C}$
according to Ref. \cite{EngineeringToolBox}). With these parameter values,
expression (\ref{4.26}) yields%
\[
\delta\approx8.8\times10^{-5},
\]
i.e., the effect of diffusion in the vdW layer is weak.

Further estimates show that diffusion is weak in the interface as well. It is
important only at a macroscopic scale, where the diffusive flux matches the
evaporative flux emitted by the vdW layer. This part of the setting, however,
is not at issue in the present paper.

To take advantage of the smallness of $\delta$, observe that the limit
$\delta\rightarrow0$ corresponds to $D\rightarrow\infty$ -- hence, Eq.
(\ref{4.19}) becomes%
\begin{equation}
\frac{\partial\hat{G}_{1}}{\partial z}-F_{1}-\frac{\partial\hat{G}_{2}%
}{\partial z}+F_{2}=0. \label{4.27}%
\end{equation}
Eqs. (\ref{4.16})--(\ref{4.17}), (\ref{4.27}) and boundary conditions
(\ref{2.34})--(\ref{2.35}), (\ref{4.22})--(\ref{4.23}) form a boundary-value
problem for the unknowns $\rho_{1}$, $\rho_{2}$, and $w$. The diffusive flux
$J$ and Eq. (\ref{4.15}) decouple from the other unknowns and equations --
thus, can be omitted.

\section{Evaporation of a liquid into air: the solution\label{sec5}}

Let%
\[
\rho_{i}=\rho_{i}(z_{new}),\qquad w=w(z_{new}),
\]%
\[
z_{new}=z+\frac{E}{\rho_{1}^{(l)}+\rho_{2}^{(l)}}t.
\]
Substituting this ansatz into Eq. (\ref{4.16}), taking into account boundary
conditions (\ref{2.34}) and (\ref{4.22}), and omitting the subscript $_{new}$,
one can deduce that%
\[
w=E\left(  \frac{1}{\rho_{1}^{(l)}+\rho_{2}^{(l)}}-\frac{1}{\rho_{1}+\rho_{2}%
}\right)  .
\]
Substituting this expression into Eqs. (\ref{4.21}) and (\ref{4.27}), one
obtains\begin{widetext}%
\begin{align*}
\rho_{1}\left(  \frac{\partial G_{1}}{\partial z}-F_{1}\right)  +\rho
_{2}\left(  \frac{\partial G_{2}}{\partial z}-F_{2}\right)   &  =E\frac
{\partial}{\partial z}\left[  \frac{\mu}{\left(  \rho_{1}+\rho_{2}\right)
^{2}}\frac{\partial\left(  \rho_{1}+\rho_{2}\right)  }{\partial z}\right]  ,\\
\left(  \frac{\partial G_{1}}{\partial z}-F_{1}\right)  -\left(
\frac{\partial G_{2}}{\partial z}-F_{2}\right)   &  =0.
\end{align*}
These equations can be viewed as a linear set for the expressions in
parentheses on their left-hand sides; solving this set and recalling
expressions (\ref{4.18}) for $F_{i}$, one obtains%
\begin{equation}
\frac{\partial}{\partial z}\left[  \hat{G}_{i}-\int\sum_{j}\Psi_{ij}%
(z-z^{\prime})\rho_{j}(z^{\prime},t)\mathrm{d}z^{\prime}\right]  =-\frac
{E}{\rho_{1}+\rho_{2}}\frac{\partial}{\partial z}\left[  \frac{\mu}{\left(
\rho_{1}+\rho_{2}\right)  ^{2}}\frac{\partial\left(  \rho_{1}+\rho_{2}\right)
}{\partial z}\right]  .\label{5.1}%
\end{equation}
\end{widetext}Since the air density is much smaller than the liquid density,
this equation can be simplified the same way the equilibrium problem was: Eq.
(\ref{5.1})$_{i=1}$ can be replaced with its pure-fluid equilibrium version,
and in Eq. (\ref{5.1})$_{i=2}$, one can set $\rho_{1}=\rho_{1}^{(sat)}$ and
replace $\hat{G}_{i}$ with its low-density asymptotic (\ref{4.13}).

Eq. (\ref{5.1}) and its asymptotic version were solved numerically, and
typical results are illustrated in Fig. \ref{fig7} together with the
corresponding results obtained via the DIM in Ref. \cite{Benilov23d}. The
following features are to be observed:

\begin{figure}
\begin{center}\includegraphics[width=\columnwidth]{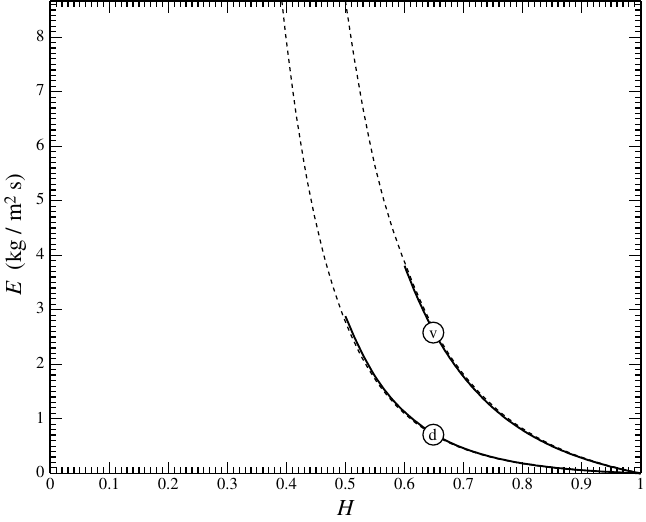}\end{center}
\caption{The evaporation rate $E$ vs. $H$, for $T=25^{\circ }\mathrm{C}$. The solid line shows the numerical solution of the exact equations and the dotted line, the asymptotic result for the limit $\rho _{2}^{(a.sat)}\ll\rho_{1}^{(l.sat)}$. Curves (v) and (d) are computed using the DIM and VM, respectively.}
\label{fig7}
\end{figure}

\begin{itemize}
\item The dependence of the evaporation rate on the relative humidity is
strongly nonlinear (unlike that for evaporation of a pure liquid into its vapor).

\item The DIM noticeably underestimates the evaporation rate (similarly to the
case of pure liquids).
\end{itemize}

It is instructive to compare the absolute value of $E$ for the two kinds of
evaporation. Using in both cases the Vlasov model, one obtains for water at
$T=25^{\circ}\mathrm{C}$ and $H=0.5$%
\begin{align*}
E  &  \approx2.0\times10^{-5}\,\mathrm{kg\,m}^{-2}\mathrm{s}^{-1}%
\qquad\text{(liquid}\rightarrow\text{vapor),}\\
E  &  \approx8.7\times10^{+1}\,\mathrm{kg\,m}^{-2}\mathrm{s}^{-1}%
\qquad\text{(liquid}\rightarrow\text{air).}%
\end{align*}
The huge difference between the two kinds of evaporation is due to the fact
that, at $25^{\circ}\mathrm{C}$, air is much denser than water vapor -- as a
result, the former exerts on the evaporating molecules a much stronger vdW
force than the latter. This force pulls the molecules forward -- hence, helps
evaporation. The back-pulling force exerted by the liquid is the same in both
cases, plus it is countered by the pressure gradient. Thus, the net force
exerted in the liquid/air system is much more conducive for evaporation than
that in the liquid/vapor system.

Note that Eq. (\ref{5.1}) is much more difficult to solve numerically than its
pure-fluid counterpart, for both DIM and VM. The difficulty is probably caused
by the presence of \emph{two} small parameters: the vapor-to-air and
vapor-to-liquid density ratios -- whereas the pure-fluid problem involves only
the latter. As a result, it was impossible to extend the curves in Fig.
\ref{fig7} to $H\lesssim0.5$. For equilibrium interfaces, the difficulties are
not as severe (which was the case with pure fluids as well), and there are no
difficulties whatsoever for the asymptotic version of the DIM (reduced to an
algebraic equation in Ref. \cite{Benilov23d}).

\section{How can the new effect be observed/simulated and why has this not
happened already?\label{sec6}}

It remains to discuss why the shortcomings of the HKL claimed in the present
paper have not been so far noticed by experimental and molecular-dynamics communities.

\subsection{Experiments\label{sec6.1}}

(i) The author of the present paper found several experimental studies of
liquids evaporating into their vapor -- but only in a forced setting, where
the vapor is sucked out of the container by a pump (e.g., Refs.
\cite{SodtkeAjaevStephan08,KazemiNobesElliott18}). The low pressure created by
the pump accelerates the evaporation and makes the predictions of the present
paper inapplicable.

More generally, the experimental community do not seem to be concerned with
unforced evaporation into \emph{vapor}, believing that it is similar to the
unforced evaporation into \emph{air} -- thus, \textquotedblleft why would
someone go to significant trouble and expense to do [such] experiments [...]
when these can be done in ambient air?\textquotedblright\ (a private
communication from Janet Elliott, Canada Research Chair in
Thermodynamics).\smallskip

(ii) As for liquids evaporating into air, the available measurements of the
coefficient $\theta$ vary between $0.01$ and $1$ for the same temperature
\cite{EamesMarrSabir97,MarekStraub01}, indicating a problem in the functional
dependence where this coefficient appears.

In other words, the experimental community do seem to have noticed the HKL's
second shortcoming claimed in this paper.\smallskip

With this said, experiments with both kinds of evaporation are objectively
difficult (hence, potentially inaccurate) because the measurements have to be
carried out very near the interface, but without interfering with the
evaporative flow.

\subsection{Molecular dynamics\label{sec6.2}}

There is a significant body of work where the fluid is approximated by a large
set of particles interacting through a potential $\phi(r)$ involving both
repulsive and attractive components. This approach, usually referred to as
molecular dynamics, has been applied to evaporation, and recent papers
\cite{LiangBibenKeblinski17,BirdGutierrezplascenciaKeblinskiLiang22} claim
that the HKL holds with an order-one $\theta$. In the present paper, on the
other hand, such is observed only for evaporation of a liquid into its vapor,
and only at a mid-range $T$ (see Fig. \ref{fig5}).

Unfortunately, a meaningful comparison between molecular dynamics and Vlasov
model is impossible at this stage.

To understand why, note that the choice of the potential $\phi(r)$ fixes all
of the fluid's characteristics, and some of them do not necessarily match the
fluid under consideration. The full match can only be accidental, in fact, as
none of the commonly-used potentials involves enough adjustable constants to
cover the parameter space of a `real' fluid. As a result, the region where the
HKL does not hold could have simply been missed.

Indeed, consider the Lennard-Jones (LJ) potential,%
\[
\phi(r)=4\epsilon\left[  \left(  \frac{r_{0}}{r}\right)  ^{12}-\left(
\frac{r_{0}}{r}\right)  ^{6}\right]  ,
\]
used in Ref. \cite{LiangBibenKeblinski17} with $r_{0}=3.41\,\mathrm{\mathring
{A}}$ and $\epsilon=10.3\,\mathrm{meV}$, to approximate argon. The triple and
critical temperatures corresponding to this $\phi(r)$ are
\cite{StephanHasse19}%
\[
T_{tr}\approx79\,\mathrm{K},\qquad T_{cr}\approx158\,\mathrm{K},
\]
whereas those of the `real' argon are \cite{AngusArmstrongGosmanEtal72}%
\[
T_{tr}\approx88\,\mathrm{K},\qquad T_{cr}\approx151\,\mathrm{K}.
\]
Unfortunately, such discrepancies are inevitable, as the LJ potential allows
one to explore only a \emph{two}-dimensional surface (parameterized by
$\epsilon$ and $r_{0}$) in the problem's \emph{multi}dimensional parameter space.

Furthermore, Ref. \cite{LiangBibenKeblinski17} employed a \emph{truncated} LJ
potential ($\phi=0$ for $r>3.2\,r_{0}$), and the effect of truncation on the
fluid's properties is difficult to assess. For example, it can be the reason
why liquid/vapor interfaces were observed in Ref. \cite{LiangBibenKeblinski17}
at a lower temperature ($T=76.3\,\mathrm{K}$) than both of the above values of
$T_{tr}$.

The mismatch of capillary characteristics is even larger than that of the
thermodynamic ones: the LJ value of argon's surface tension -- say, at the
triple point -- is $\gamma_{tr}=18.6\,\mathrm{dyn/cm}$ \cite{StephanHasse19},
whereas its `real' value is $\gamma_{tr}=12.6\,\mathrm{dyn/cm}$
\cite{Stansfield58}.

Finally, the vapor-to-liquid density ratio corresponding to the truncated LJ
potentials can be very different from that of the `real' fluid -- and this is
the most important mismatch of all.

In the simulations of water evaporating into ambient nitrogen reported in Ref.
\cite{BirdGutierrezplascenciaKeblinskiLiang22}, this parameter was
\begin{equation}
\frac{\rho_{1}^{(a.sat)}}{\rho_{1}^{(l.sat)}}\gtrsim2.1\times10^{-3},
\label{6.1}%
\end{equation}
whereas, for `real' water at, say, $25^{\circ}\mathrm{C}$, the vapor-to-liquid
density ratio is smaller by two orders of magnitude,%
\[
\frac{\rho_{1}^{(v.sat)}}{\rho_{1}^{(l.sat)}}\approx2.3\times10^{-5}.
\]
Furthermore, Ref. \cite{BirdGutierrezplascenciaKeblinskiLiang22} simulated the
case where the vapor and ambient-gas densities were comparable,%
\[
\frac{\rho_{1}^{(a.sat)}}{\rho_{2}^{(a.sat)}}\sim0.7\div0.9,
\]
whereas, in the `real' atmosphere at $25^{\circ}\mathrm{C}$, this parameter is
small,%
\[
\frac{\rho_{1}^{(a.sat)}}{\rho_{2}^{(a.sat)}}\approx1.9\times10^{-2}.
\]
Since the intermolecular force in the vdW layer crucially depends on the
density and composition of air, the differences in these characteristics
explain the disagreement between the present results and those of Ref.
\cite{BirdGutierrezplascenciaKeblinskiLiang22}.

More generally, to reconcile the Vlasov model and molecular dynamics, one should

\begin{itemize}
\item either use molecular dynamics with a potential $\phi(r)$ involving
enough adjustable parameters to mimic a common liquid under normal conditions;

\item or apply the Vlasov model to a fluid whose characteristics match those
of the truncated LJ potential.
\end{itemize}

In the context of the latter approach, note that, for pure water, condition
(\ref{6.1}) holds if $T\gtrsim155^{\circ}\mathrm{C}$ -- and the corresponding
values of $\theta$ computed via the Vlasov model are order one (see Fig.
\ref{fig5}). One can further assume that a small proportion of ambient
nitrogen (as in Ref. \cite{BirdGutierrezplascenciaKeblinskiLiang22}) should
not alter the VM results too strongly.

\section{Summary and concluding remarks\label{sec7}}

This work examines the effect of the van der Waals force on evaporation. The
following conclusions have been drawn:

\begin{enumerate}
\item[(i)] For evaporation of a liquid into its vapor, the dependence of the
evaporation rate $E$ on the relative humidity $H$ is almost linear -- hence,
the Hertz--Knudsen Law (HKL) is functionally correct. Yet the
evaporation/condensation probability $\theta$, which appears as a coefficient
in the HKL, is much smaller than unity, making evaporation much slower than expected.

\item[(ii)] For evaporation of a liquid into air, the dependence of $E$ on $H$
is strongly nonlinear, so the HKL does not seem to apply functionally.
\end{enumerate}

\noindent Conclusion (i)--(ii) are illustrated by Figs. \ref{fig5} and
\ref{fig7}, respectively.

In addition to physical conclusions, a technical one has been drawn, which
might be important for researchers utilizing the diffuse-interface model:

\begin{enumerate}
\item[(iii)] The DIM fails in a certain region (the vdW layer) at the
outskirts of the interface and, as a result, noticeably underestimates the
evaporative flux by comparison with the more accurate Vlasov model.
\end{enumerate}

\noindent It should be noted, however, that even though the DIM comes short in
application to evaporation, it remains to be seen whether it does so in other
settings (contact lines, cavitation, etc.). It all depends on whether the DIM
solution involves a short-scale boundary layer (vdW layer), making it
inapplicable. Furthermore, conclusion (iii) does not apply to a whole class of
DIM models -- those where the density in the interfacial region changes
gradually, but the vdW force is not included (e.g.,
\cite{ChiapolinoBoivinSaurel16,ChiapolinoBoivinSaurel17,DengBoivin20}).

One should also keep in mind that all conclusions of this work have been drawn
using the \emph{hydrodynamic} approximation of evaporation, which does not
describe \emph{kinetic} effects -- such as, for example, the temperature jump
associated with the Knudsen layer
\cite{BondStruchtrup04,RanaLockerbySprittles18}.

To compare the kinetic effects to that of the vdW force, one needs to switch
to a kinetic model -- e.g., the Enskog--Vlasov equation
\cite{Desobrino67,Grmela71,GrmelaGarciacolin80,GrmelaGarciaColin80b,FrezzottiGibelliLorenzani05,BarbanteFrezzottiGibelli15,FrezzottiBarbante17,FrezzottiGibelliLockerbySprittles18,BenilovBenilov18,BenilovBenilov19a,BenilovBenilov19b}%
. This is what the author of the present paper initially intended to do, but
such a large increase in the model's complexity turned out to be
insurmountable in a single stride.

As an alternative to the Enskog--Vlasov kinetic equation, one might use the
multi-moment model derived in Ref. \cite{StruchtrupFrezzotti22}. It has a
better chance of yielding a relatively simple expression for the evaporative
flux, suitable for the use in natural, biological, and industrial applications.

\acknowledgements{I am grateful to Janet Elliott and Zhi Liang for helping me to understand their work.}

\appendix

\section{The parameters used in the paper\label{appA}}

This appendix describes how the parameters involved in the DIM and VM can be
determined for a specific fluid. The examples considered are water and air;
the latter is treated as a mixture of nitrogen and oxygen.

\subsection{The parameters of the Enskog--Vlasov fluid model\label{appA.1}}

The results described in this subsection were originally reported in Ref.
\cite{Benilov23a,Benilov23d} and are presented here for completeness.

\subsubsection{The van der Waals parameter of a pure fluid\label{appA.1.1}}

To determine the vdW parameter $a$ for a pure fluid, observe that the
Enskog--Vlasov expression (\ref{2.7}) for the internal energy $e(\rho,T)$ is
linear in $\rho$. This allows one to determine $a$ as the slope of a linear
fit to the empiric dependence of $cT-e$ on $\rho$, where the heat capacity $c$
is the same as that in the Enskog--Vlasov (kinetic) theory -- i.e., $3R$ for
water and $5R/2$ for nitrogen and oxygen. For simplicity, the fitting was
carried out using only the data on the critical isobar $p=p_{cr}$, but the
resulting straight line fits the isobars $p=p_{cr}/2$ and $p=2p_{cr}$
reasonably well too (see Fig. 9(a) of Ref. \cite{Benilov23a}).

The values of $a$ determined this way for water, nitrogen, and oxygen are
listed in Table \ref{tab2}. It also includes the van der Waals parameter of
air (calculated as the 79/21 weighted average of those of nitrogen and oxygen, respectively).

\begin{table*}
\renewcommand{\arraystretch}{1.6}\centering
\begin{ruledtabular}
\caption{The parameters of $\mathrm{H}_{2}\mathrm{O}$, $\mathrm{N}_{2}$, $\mathrm{O}_{2}$, and air: $R$ is the specific gas constant, $a$ is the van der Waals parameter, $K_{D}$ and $K_{V}$ are the values of the Korteweg parameter according to the DIM and VM, respectively. The parameters of air are calculated as the 79/21 weighted averages of the corresponding parameters of nitrogen and oxygen, respectively.\vspace{2mm}}
\begin{tabularx}{\textwidth}{lcccc}
Fluid & $R~(\mathrm{m}^{2}\mathrm{s}^{-2}\mathrm{K}^{-1})$ & $a~(\mathrm{m}^{5}\mathrm{s}^{-2}\mathrm{kg}^{-1})$ &
$K_{D}~(\mathrm{m}^{7}\mathrm{s}^{-2}\mathrm{kg}^{-1})$ & $K_{V}~(\mathrm{m}^{7}\mathrm{s}^{-2}\mathrm{kg}^{-1})$ \\[5pt]
\hline
$\mathrm{H}_{2}\mathrm{O}$ & $461.52$ & $2112.1$ & $1.8781\times 10^{-17}$ & $2.2906\times 10^{-17}$ \\
$\mathrm{N}_{2}$ & $296.81$ & $222.2$ & $1.5078\times 10^{-17}$ & $1.6998\times 10^{-17}$ \\
$\mathrm{O}_{2}$ & $259.84$ & $172.7$ & $0.8459\times 10^{-17}$ & $1.0203\times 10^{-17}$ \\
air & $289.05$ & $211.8$ & $1.3688\times 10^{-17}$ & $1.5571\times 10^{-17}$ \\
\end{tabularx}
\label{tab2}
\end{ruledtabular}
\end{table*}

\subsubsection{Thermodynamic properties of a pure fluid\label{appA.1.2}}

According to the Enskog--Vlasov fluid model (\ref{2.8})--(\ref{2.9}), the
properties of a pure fluid are described by its van der Waals parameter $a$
(see the previous subsubsection) and the non-ideal part of the entropy,
$\Theta(\rho)$. The latter should include enough undetermined coefficients to
fit the fluid's empiric equation of state. The following expression was
suggested in Ref. \cite{Benilov23a}:%
\begin{multline}
\Theta(\rho)=-Rq^{(0)}\ln\left(  1-0.99\frac{\rho}{\rho_{tp}}\right) \\
+R\sum_{n=1}^{4}q^{(n)}\left(  \frac{\rho}{\rho_{tp}}\right)  ^{n},
\label{A.1}%
\end{multline}
where $q^{(0)}$... $q^{(4)}$ are undetermined coefficients and $\rho_{tp}$ is
the fluid's density at the triple point ($\rho_{tp}$ is simply a convenient
density scale; the fact that, at the triple point, all three phases are in
equilibrium is irrelevant).

The coefficients $q^{(n)}$ were determined by ensuring that the expressions
for $p(\rho,T)$ and $G(\rho,T)$ corresponding to (\ref{A.1}) yield the
`correct' -- i.e., empiric -- values for the critical density, temperature,
and pressure, as well as the liquid and vapor densities at the triple point
(five equations for the five unknown coefficients).

The values of $q^{(n)}$ for water as determined in Ref. \cite{Benilov23a} are%
\[
q^{(0)}=0.071894,\qquad q^{(1)}=1.4139,\qquad q^{(2)}=8.1126,
\]%
\[
q^{(3)}=-8.3669,\qquad q^{(4)}=4.0238,
\]
and those for nitrogen and oxygen can be found in Table 3 of the same paper.
The accuracy with which the resulting equations of state approximate the
empiric ones is illustrated in Fig. 9(c) of Ref. \cite{Benilov23a}.

\subsubsection{The van der Waals matrix of a binary mixture\label{appA.1.3}}

When modeling evaporation of water into air, one needs $a_{11}$ (water--water
interaction), $a_{22}$ (air--air interaction), and $a_{12}$ (water--air
interaction). The first two can be found in Table \ref{tab2}, and $a_{12}$ can
be deduced from a single measurement of the density of air dissolved in water
at a certain temperature and pressure.

To do so, consider the equilibrium interface, so that $\rho_{2}^{(l.sat)}$ is
the density of the air dissolved in water. It generally depends on $a_{12}$ --
which can, thus, be fixed by fitting $\rho_{2}^{(l.sat)}$ to its empiric
value. For water at $T=25^{\circ}\mathrm{C}$ and $p=1\,\mathrm{atm}$, for
example, Ref. \cite{TheEngineeringToolbox-AirWaterSolubility} yields $\rho
_{2}^{(l.sat)}=0.0227\,\mathrm{kg\,m}^{-3}$, and this value emerges from the
Maxwell construction (\ref{4.5})--(\ref{4.8}) only if%
\[
a_{12}=208.2\,\mathrm{m}^{5}\mathrm{s}^{-2}\mathrm{kg}^{-1}.
\]
Note that this value is specific to the Enskog--Vlasov fluid model used with
the Maxwell construction.

\subsection{The Korteweg parameter\label{appA.2}}

A fluid's thermodynamic properties (discussed above) do not depend on the
chosen model of the vdW force, but the \emph{capillary} properties do. As a
result, the DIM and VM correspond to different values of the Korteweg
parameter, which will be denoted by $K_{D}$ and $K_{V}$, respectively.

\subsubsection{The Korteweg parameter of a pure fluid\label{appA.2.1}}

A fluid's Korteweg parameter can be deduced from a single measurement of its
surface tension -- say, at the triple point. For water, nitrogen, and oxygen,
such measurements can be found in Ref. \cite{Somayajulu88}.

Consider a flat equilibrium interface described by its density profile
$\rho^{(sat)}(z)$. Then, according to the DIM, the surface tension is
\cite{PismenPomeau00}%
\begin{equation}
\gamma_{D}=K_{D}\int\left(  \frac{\mathrm{d}\rho^{(sat)}}{\mathrm{d}z}\right)
^{2}\mathrm{d}z. \label{A.2}%
\end{equation}
In Ref. \cite{Benilov23a}, $\gamma_{D}$ was calculated for water, nitrogen,
and oxygen at their respective triple points -- and the values of $K_{D}$ were
determined, such that (\ref{A.2}) agrees with the corresponding empiric
result. These values of $K_{D}$ are listed in Table \ref{tab2}.

To find $K_{V}$, one should first derive the Vlasov equivalent of formula
(\ref{A.2}). To do so, consider a static macroscopic drop of radius $R$ and
calculate the pressure difference between the inside and outside. One should
expect it to be of the form $\gamma_{V}/R$, where the coefficient $\gamma_{V}$
is the desired surface tension.

A static ($w=0$) density distribution $\rho(\mathbf{r})$ is described by the
following reduction of the momentum equation:%
\begin{equation}
\mathbf{\nabla}\hat{p}(\rho,T)=\rho\mathbf{\nabla}\int\rho(\mathbf{r}^{\prime
})\,\Phi(\left\vert \mathbf{r}-\mathbf{r}^{\prime}\right\vert )\,\mathrm{d}%
^{3}\mathbf{r}^{\prime}. \label{A.3}%
\end{equation}
Write the integral on the right-hand side of (\ref{A.3}) in spherical
coordinates $\left(  r^{\prime},\beta,\alpha\right)  $ and let the azimuthal
angle $\alpha$ be measured from the direction of $\mathbf{r}$. Then, for a
spherically symmetric $\rho$, the integrand does not depend on the polar angle
$\beta$, and one can reduce (\ref{A.3}) to%
\begin{equation}
\frac{\mathrm{d}\hat{p}(\rho,T)}{\mathrm{d}r}=\rho\frac{\mathrm{d}}%
{\mathrm{d}r}\int_{0}^{\infty}\rho(r^{\prime})\,\Omega(r,r^{\prime
})\,r^{\prime2}\mathrm{d}r^{\prime}, \label{A.4}%
\end{equation}
where%
\begin{equation}
\Omega(r,r^{\prime})=2\pi\int_{0}^{\pi}\Phi\left(  \sqrt{r^{2}+r^{\prime
2}-2rr^{\prime}\cos\alpha}\right)  \sin\alpha\,\mathrm{d}\alpha. \label{A.5}%
\end{equation}
Since the intermolecular potential $\Phi(r)$ decays as $r\rightarrow\infty$,
one can show that the function $\Omega(r,r^{\prime})$ decays as $\left\vert
r-r^{\prime}\right\vert \rightarrow\infty$. One can also verify (see Appendix
\ref{appF}) that relationship (\ref{4.24}) between $\Phi$ and $a$ implies that%
\begin{equation}
\int_{r}^{\infty}\Omega(r,r^{\prime})\,r^{\prime2}\mathrm{d}r^{\prime
}\rightarrow-a\qquad\text{as}\qquad r\rightarrow\infty. \label{A.6}%
\end{equation}
Impose the following boundary conditions%
\begin{equation}
\rho(r)\rightarrow\rho_{\infty}\qquad\text{as}\qquad r\rightarrow\infty,
\label{A.7}%
\end{equation}%
\[
\frac{\mathrm{d}\rho}{\mathrm{d}r}\rightarrow0\qquad\text{at}\qquad r=0.
\]
Assuming that $\rho(r)$ describes a spherical drop, introduce the liquid
density at its center,%
\begin{equation}
\rho_{0}=\rho(0), \label{A.8}%
\end{equation}
and define the drop's radius $R$ by, say,%
\[
\rho(R)=\frac{1}{2}\left(  \rho_{0}+\rho_{\infty}\right)  .
\]
Next, pick $R_{0}$ and $R_{\infty}$ such that $R_{0}<R<R_{\infty}$ and
integrate Eq. (\ref{A.4}) from $R_{0}$ to $R_{\infty}$, which yields%
\begin{equation}
p(\rho(R_{\infty}),T)-p(\rho(R_{0}),T)=I_{1}+I_{2}+I_{3}, \label{A.9}%
\end{equation}
where%
\begin{align*}
I_{1}  &  =\int_{R_{0}}^{R_{\infty}}\int_{R_{\infty}}^{\infty}\rho
(r)\,\rho(r^{\prime})\frac{\partial\Omega(r,r^{\prime})}{\partial r}%
r^{\prime2}\mathrm{d}r^{\prime}\mathrm{d}r,\\
I_{2}  &  =\int_{R_{0}}^{R_{\infty}}\int_{R_{0}}^{R_{\infty}}\rho
(r)\,\rho(r^{\prime})\frac{\partial\Omega(r,r^{\prime})}{\partial r}%
r^{\prime2}\mathrm{d}r^{\prime}\mathrm{d}r,\\
I_{3}  &  =\int_{R_{0}}^{R_{\infty}}\int_{0}^{R_{0}}\rho(r)\,\rho(r^{\prime
})\frac{\partial\Omega(r,r^{\prime})}{\partial r}r^{\prime2}\mathrm{d}%
r^{\prime}\mathrm{d}r.
\end{align*}
For a macroscopic drop -- such that $R$ is much larger than the interfacial
width $l$ -- these integrals can be simplified.

Let $R-R_{0}$ and $R_{\infty}-R$ be much larger than $l$. Given boundary
condition (\ref{A.7}), one can in $I_{1}$ set $\rho(r)\approx\rho(r^{\prime
})\approx\rho_{\infty}$ and obtain%
\[
I_{1}\approx\rho_{\infty}^{2}\int_{R_{\infty}}^{\infty}\left[  \Omega
(R_{\infty},r^{\prime})-\Omega(R_{0},r^{\prime})\right]  r^{\prime2}%
\mathrm{d}r^{\prime}.
\]
Since $r^{\prime}\in\left(  R_{\infty},\infty\right)  $, the second term in
the square brackets is negligible, after which the integral can be estimated
via property (\ref{A.6}),%
\begin{equation}
I_{1}\approx-a\rho_{\infty}^{2}. \label{A.10}%
\end{equation}
In a similar fashion, one obtains%
\begin{equation}
I_{3}\approx a\rho_{0}^{2}. \label{A.11}%
\end{equation}
Before calculating $I_{2}$, one should symmetrize it with respect to $r$ and
$r^{\prime}$, which yields, after straightforward algebra,\begin{widetext}%
\[
I_{2}=\pi\int_{R_{0}}^{R_{\infty}}\int_{R_{0}}^{R_{\infty}}\rho(r)\,\rho
(r^{\prime})\int_{0}^{\pi}\left[  \frac{\mathrm{d}\Phi(\xi)}{\mathrm{d}\xi
}\right]  _{\xi=\sqrt{r^{2}+r^{\prime2}-2rr^{\prime}\cos\alpha}}\frac{\left(
r+r^{\prime}\right)  rr^{\prime}-\left(  r^{3}+r^{\prime3}\right)  \cos\alpha
}{\sqrt{r^{2}+r^{\prime2}-2rr^{\prime}\cos\alpha}}\sin\alpha\,\mathrm{d}%
\alpha\,\mathrm{d}r^{\prime}\mathrm{d}r.
\]
\end{widetext}The main contribution to this triple integral comes from the
region%
\[
r,r^{\prime}\rightarrow\infty,\qquad\left\vert r-r_{1}\right\vert
=\mathcal{O}(1),\qquad\alpha\rightarrow0,
\]
where the solution $\rho(r)$ can be approximated by that for a flat
equilibrium interface with its `midpoint' pinned to $r=R$,%
\[
\rho(r)\approx\rho^{(sat)}(r-R).
\]
Now, take the limit%
\[
R_{0}\rightarrow-\infty,\qquad R_{\infty}\rightarrow+\infty,
\]
expand $I_{2}$ in $1/R$, and omit the terms $\sim1/R^{2}$ and smaller.
Changing the variable of integration from $\alpha$ to $r_{\bot}=\sqrt{rr_{1}%
}\alpha$, one obtains (after straightforward algebra, involving integration by
parts)%
\begin{equation}
I_{2}\approx\frac{\gamma_{V}}{R}, \label{A.12}%
\end{equation}
where%
\begin{equation}
\gamma_{V}=\int\int\rho^{(sat)}(z)\,\rho^{(sat)}(z^{\prime})\,\chi
(z-z^{\prime})\,\mathrm{d}z^{\prime}\mathrm{d}z, \label{A.13}%
\end{equation}%
\begin{equation}
\chi(z)=\pi z^{2}\Phi(z)-\pi\int_{0}^{\infty}\Phi\left(  \sqrt{z^{2}+r_{\bot
}^{2}}\right)  r_{\bot}\mathrm{d}r_{\bot}. \label{A.14}%
\end{equation}
To ascertain that $\gamma_{V}$ is the surface tension of the Vlasov model, one
should substitute expressions (\ref{A.10})--(\ref{A.12}) into Eq. (\ref{A.9})
and use (\ref{2.11}) to express $\hat{p}$ through $p$. The resulting equality
shows that the (full) pressure difference between the inside and outside of
the drop is indeed $\gamma_{V}/R$.

Applying expressions (\ref{A.13})--(\ref{A.14}) [with $\Phi$ given by
(\ref{2.23}), (\ref{2.25})--(\ref{2.26}), (\ref{2.28})] to water, nitrogen,
and oxygen, and making sure that the results match the empiric ones from Ref.
\cite{Somayajulu88}, one obtains the values of $K_{V}$ listed in table
\ref{tab2}.

\subsubsection{The Korteweg matrix of a binary mixture\label{appA.2.2}}

When modeling evaporation of water into air, one needs $K_{11}$ (water--water
interaction), $K_{22}$ (air--air interaction), and $K_{12}$ (water--air
interaction). The first two coefficients are listed in Table \ref{tab2}, and
$K_{12}$ is discussed below.

In principle, $K_{12}$ could be determined by comparing the characteristics of
water/\emph{air} and water/\emph{water-vapor} interfaces, but the difference
between the two at normal conditions is too small to be reliably measured.
Alternatively, $K_{12}$ can be deduced from the surface tension of the
water/air interface \emph{at high pressure}, and in Ref. \cite{Benilov23d},
this was done for the DIM, using the empiric results of Ref.
\cite{HintonAlvarez21}. Unfortunately, the measurements reported in Ref.
\cite{HintonAlvarez21} are scarce (only 6 points) and a bit noisy -- thus, the
accuracy of the resulting value of $K_{12}$ was difficult to assess.

For the present work, the coefficient $K_{12}$ was derived from the properties
of water/nitrogen and water/oxygen interfaces, both at high pressure, reported
in Ref. \cite{MassoudiKing74}. The following procedure was used.

Let a flat equilibrium interface in a multicomponent fluid be characterized by
$\rho_{i}^{(sat)}(z)$ -- say, computed using the Vlasov model. Then, the
multicomponent analogue of the Vlasov expression (\ref{A.13}) for the surface
tension is%
\begin{equation}
\gamma_{V}=\sum_{i,j}\int\int\rho_{i}^{(sat)}(z)\,\rho_{j}^{(sat)}(z^{\prime
})\,\chi_{ij}(z-z^{\prime})\,\mathrm{d}z^{\prime}\mathrm{d}z, \label{A.15}%
\end{equation}
where $\chi_{ij}(z)$ is given by (\ref{A.14}) with $\Phi$ replaced with
$\Phi_{ij}$. The coefficient $K_{12}$ can be determined by fitting the above
theoretical expression to the empiric dependence of $\gamma$ on $p$.

\begin{figure}
\begin{center}\includegraphics[width=\columnwidth]{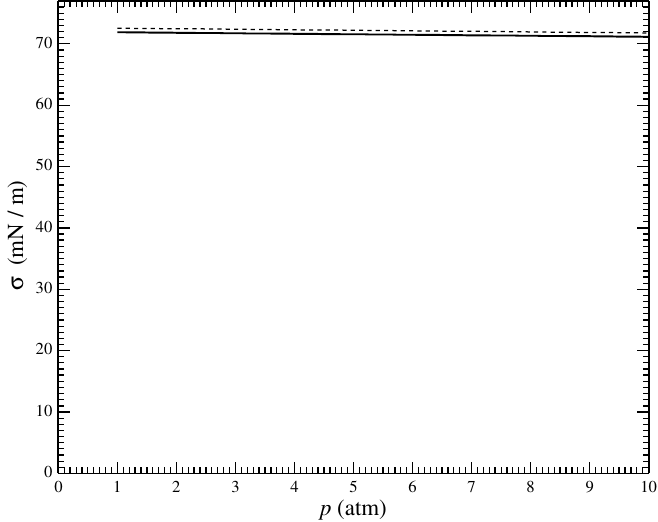}\end{center}
\caption{The surface tension of water/nitrogen interface vs. the pressure. The solid curve shows the empiric results of Ref. \cite{MassoudiKing74}, the dashed curve is computed using the Vlasov model.}
\label{fig8}
\end{figure}

An example of such a fit is presented in Fig. \ref{fig8}. One can see that the
slope of the theoretical curve is close to that of the empiric one, but the
two are separated by a gap. This is because the measurements of Ref.
\cite{MassoudiKing74} were carried out at $25^{\circ}\mathrm{C}$ -- whereas,
in the present paper, the Vlasov model is tuned to yield the correct surface
tension of pure water at $0^{\circ}\mathrm{C}$. In principle, the gap could be
eliminated by retuning the model for $25^{\circ}\mathrm{C}$, but that would
only marginally improve the overall accuracy -- hence, is not worth implementing.

The same procedure was also carried out for the DIM, in which case expression
(\ref{A.15}) should be replaced with
\[
\gamma_{D}=\sum_{i,j}K_{ijD}\int\frac{\mathrm{d}\rho_{i}^{(sat)}}{\mathrm{d}%
z}\frac{\mathrm{d}\rho_{j}^{(sat)}}{\mathrm{d}z}\mathrm{d}z.
\]
The calculated values of $K_{12D}$ and $K_{12V}$ are listed in Table
\ref{tab3}.

\begin{table}
\renewcommand{\arraystretch}{1.6}\centering
\begin{ruledtabular}
\caption{The nondiagonal Korteweg parameter $K_{12}$ of water/nitrogen, water/oxygen, and water/air interfaces. $\left( K_{12}\right)  _{D}$ and $\left(  K_{12}\right)  _{V}$ are calculated according the DIM and VM, respectively.\vspace{2mm}}
\begin{tabularx}{\columnwidth}{lcc}
Interface & $K_{12D}~(\mathrm{m}^{7}\mathrm{s}^{-2}\mathrm{kg}^{-1})$ & $K_{12V}~(\mathrm{m}^{7}\mathrm{s}^{-2}\mathrm{kg}^{-1})$ \\[5pt]
\hline
$\mathrm{H}_{2}\mathrm{O}$/$\mathrm{N}_{2}$ & $0.6816\times 10^{-17}$ & $1.8548\times 10^{-17}$\\
$\mathrm{H}_{2}\mathrm{O}$/$\mathrm{O}_{2}$ & $0.6126\times 10^{-17}$ & $1.1863\times 10^{-17}$\\
$\mathrm{H}_{2}\mathrm{O}$/air & $0.6671\times 10^{-17}$ & $1.7144\times 10^{-17}$\\
\end{tabularx}
\label{tab3}
\end{ruledtabular}
\end{table}

\subsection{The viscosity function $\mu(\rho,T)$\label{appA.3}}

When modeling evaporation of water into air, one needs the shear and bulk
viscosities, $\mu_{s}$ and $\mu_{b}$, of both \emph{air} and \emph{water
vapor}. The characteristics of \emph{liquid water} are asymptotically
unimportant -- as shown in the present work and Ref. \cite{Benilov23d} for the
VM and DIM, respectively (in both cases, provided the liquid's density exceeds
those of vapor and air).

In the present work, $\mu_{s}$ and $\mu_{b}$ of air are calculated using the
empiric formulae of Ref. \cite{ShangWuWangYangYeHuTaoHe19}; and $\mu_{s}$ of
water vapor, using the IAPWS formulae \cite{WagnerPruss02}.

As for $\mu_{b}$ of water vapor, there seems to be only one source for it --
Ref. \cite{Cramer12}. In this paper, the results are presented in graphical
form, for the interval $58^{\circ}\mathrm{C}\lessapprox T\lessapprox
651^{\circ}\mathrm{C}$. The author of the present work digitized them and
extrapolated to $T=0^{\circ}\mathrm{C}$. It is worth mentioning here that, at
normal conditions, the density of vapor is much smaller than that of air,
making the characteristics of the former asymptotically unimportant.

Finally, the effective viscosity $\mu$ of the mixture of air and vapor was
calculated using the mixture rule\ proposed in Ref.
\cite{HindMclaughlinUbbelohde60},%
\[
\mu=\mu_{1}\left(  \frac{\rho_{1}}{m_{1}}\right)  ^{2}+\left(  \mu_{1}+\mu
_{2}\right)  \frac{\rho_{1}}{m_{1}}\frac{\rho_{2}}{m_{2}}+\mu_{2}\left(
\frac{\rho_{2}}{m_{2}}\right)  ^{2},
\]
where the subscripts $1$ and $2$ correspond to the vapor and air, respectively.

\section{The isobaricity condition\label{appB}}

To prove the isobaricity condition (\ref{2.36}) for the Vlasov model,
substitute expression (\ref{2.17}) for the vdW force into the momentum
equation (\ref{2.30}), take into account that%
\[
\frac{\mathrm{d}\Psi(z-z^{\prime})}{\mathrm{d}z}=-\frac{\mathrm{d}%
\Psi(z-z^{\prime})}{\mathrm{d}z^{\prime}},
\]
and write it in the form%
\[
\frac{\partial}{\partial z}\left[  \hat{p}(\rho,T)-\mu(\rho,T)\frac{\partial
w}{\partial z}\right]  =-\rho\int\rho(z^{\prime},t)\frac{\mathrm{d}%
\Psi(z-z^{\prime})}{\mathrm{d}z^{\prime}}\mathrm{d}z^{\prime}.
\]
Integrating this equation from $-Z$ to $Z$ (where $Z>0$ is a large but finite
distance), taking the limit $Z\rightarrow\infty$, and recalling boundary
conditions (\ref{2.32})--(\ref{2.33}), one obtains%
\begin{equation}
\hat{p}(\rho^{(v)},T)-\hat{p}(\rho^{(l)},T)=-I, \label{B.1}%
\end{equation}
where%
\begin{equation}
I=\lim_{Z\rightarrow\infty}\int_{-Z}^{Z}\rho(z)\int_{-\infty}^{\infty}%
\rho(z^{\prime})\frac{\mathrm{d}\Psi(z-z^{\prime})}{\mathrm{d}z^{\prime}%
}\mathrm{d}z^{\prime}\mathrm{d}z. \label{B.2}%
\end{equation}
One might be tempted to take the limit $Z\rightarrow\infty$, and then convert
$I$ from a repeated to double integral; the latter would have symmetric limits
but antisymmetric integrand -- hence, $I=0$.

Such a calculation would be incorrect, however: the limit $Z\rightarrow\infty$
should not be taken before the conversion of $I$ to double integration, as
this is not the order in which these operations appear in (\ref{B.2}).
Furthermore, changing the order of integration is not generally allowed in
improper integrals.

Instead, rearranged $I$ in the form%
\begin{equation}
I=I_{1}+I_{2}+I_{3}, \label{B.3}%
\end{equation}
where%
\begin{align*}
I_{1}  &  =\lim_{Z\rightarrow\infty}\int_{-Z}^{Z}\rho(z)\int_{Z}^{+\infty}%
\rho(z^{\prime})\frac{\mathrm{d}\Psi(z-z^{\prime})}{\mathrm{d}z^{\prime}%
}\mathrm{d}z^{\prime}\mathrm{d}z,\\
I_{2}  &  =\lim_{Z\rightarrow\infty}\int_{-Z}^{Z}\rho(z)\int_{-Z}^{Z}%
\rho(z^{\prime})\frac{\mathrm{d}\Psi(z-z^{\prime})}{\mathrm{d}z^{\prime}%
}\mathrm{d}z^{\prime}\mathrm{d}z,\\
I_{3}  &  =\lim_{Z\rightarrow\infty}\int_{-Z}^{Z}\rho(z)\int_{-\infty}%
^{-Z}\rho(z^{\prime})\frac{\mathrm{d}\Psi(z-z^{\prime})}{\mathrm{d}z^{\prime}%
}\mathrm{d}z^{\prime}\mathrm{d}z.
\end{align*}
The region of integration in $I_{2}$ is finite -- hence, $I_{2}$ can be safely
converted to double integration (with symmetric limits and antisymmetric
integrand) -- hence, $I_{2}=0$.

To evaluate $I_{1}$, observe that%
\[
z<Z<z^{\prime},
\]
and keep in mind that $z$ and $z^{\prime}$ cannot be wide apart -- otherwise
the contribution of such a pair to the integral would be negligible, because%
\[
\frac{\mathrm{d}\Psi(z-z^{\prime})}{\mathrm{d}z}\rightarrow0\qquad
\text{as}\qquad\left\vert z-z^{\prime}\right\vert \rightarrow\infty.
\]
This means that, as $Z\rightarrow\infty$, both $\rho(z)$ and $\rho(z^{\prime
})$ in $I_{1}$ can be replaced with $\rho^{(v)}$ -- and recalling definition
(\ref{2.20}) of $a$, one obtains $I_{2}=a\rho^{(v)2}$.

Similarly, one obtains $I_{3}=-a\rho^{(l)2}$, after which Eqs. (\ref{B.1}),
(\ref{B.3}), and relationship (\ref{2.11}) between $p$ and $\hat{p}$ yield the
isobaricity condition (\ref{2.36}), as required.

\section{The numerical method\label{appC}}

When solving equation Eq. (\ref{3.2}), the unknown $\rho(z)$ was discretized
on a uniform mesh. Outside the computational region, $\rho(z)$ was equated to
either $\rho^{(l)}$ or $\rho^{(v)}$, according to boundary conditions
(\ref{2.32})--(\ref{2.33}).

The integral in Eq. (\ref{3.2}) was evaluated using the method of rectangles.
To improve its accuracy, the mesh was chosen so that the endpoints of the
integration interval are nodes -- this choice, plus the fact that $\Psi(z)$
vanishes at the endpoints, and does so with zero derivative, reduces the error
of the method of rectangles to $\mathcal{O}(step^{4})$, which is the same as
that of Simpson's rule.

The resulting set of nonlinear algebraic equations was solved using the
function FSOLVE of MATLAB.

It has turned out, however, that the above algorithm does not work for Eq.
(\ref{3.2}) in its original form: the iterations just would not converge for
all solutions except the equilibrium one (where $E=0$). This suggested that
the problem was caused by the derivative term on the right-hand side of
(\ref{3.2}) where $E$ appears as a coefficient. Various finite-difference
approximations of this term were tested, but none worked.

Eventually, it was established by trial and error that the proposed algorithm
works only if Eq. (\ref{3.2}) is first integrated with respect to $z$ from
$-\infty$ to $z^{\prime\prime}$, and then boundary condition (\ref{2.32}) is
used to obtain%
\begin{multline}
\hat{G}(\rho(z^{\prime\prime}),T)-\int\rho(z^{\prime})\,\Psi(z^{\prime\prime
}-z^{\prime})\,\mathrm{d}z^{\prime}-G(\rho^{(l)},T)\\
=-\int_{-\infty}^{z^{\prime\prime}}\frac{E}{\rho(z)}\frac{\mathrm{d}%
}{\mathrm{d}z}\left[  \frac{\mu(\rho(z),T)}{\rho^{2}}\frac{\mathrm{d}\rho
(z)}{\mathrm{d}z}\right]  \mathrm{d}z. \label{C.1}%
\end{multline}
The integral on the right-hand side of this equation was evaluated using
Simpson's rule [to make the error of the computation consistent with that of
the integral on the left-hand side of Eq. (\ref{C.1})].

\section{Evaporation of a liquid into its vapor: the $T\rightarrow0$ limit of
the DIM\label{appD}}

Evaporation of a pure fluid under the diffuse-interface approximation has been
examined in Refs. \cite{Benilov23c,Benilov23d} using a certain shortcut. In
what follows, this shortcut will be reformulated in terms of the standard
matched asymptotics, allowing one to estimate the spatial scale of the flow.

The problem will be nodimensionalized using the spatial scale $l_{F}$ of the
vdW force given by (\ref{2.22}), and the following velocity scale%
\[
\bar{w}=\frac{\bar{p}l_{F}}{\bar{\mu}},
\]
where $\bar{p}$ and $\bar{\mu}$ are the pressure and viscosity scales,
respectively. Physically, the above choice of $\bar{w}$ corresponds to the
most general regime where the pressure gradient, viscous stress, and vdW force
are of the same order \cite{Benilov23a,Benilov23d}.

Let the density scale $\bar{\rho}$ be that of liquid and set%
\[
\bar{p}=a\bar{\rho}^{2},
\]
which reflects the non-ideal part of the pressure [i.e., the second term in
the Enskog--Vlasov equation of state (\ref{2.8})].

Define the following nondimensional variables:%
\[
z_{nd}=\frac{z}{l_{F}},\qquad\rho_{nd}=\frac{\rho}{\bar{\rho}},\qquad
E_{nd}=\frac{E}{\bar{\rho}\bar{w}},
\]%
\[
\mu_{nd}=\frac{\mu}{\bar{\mu}},\qquad T_{nd}=\frac{RT}{a\bar{\rho}},
\]%
\[
p_{nd}=\frac{p}{a\bar{\rho}^{2}},\qquad G_{nd}=\frac{G}{a\bar{\rho}}.
\]
Nondimensionalizing Eq. (\ref{3.5}) and omitting the subscript $_{nd}$, one
obtains%
\begin{equation}
\frac{\mathrm{d}}{\mathrm{d}z}\left[  G(\rho,T)-\frac{\mathrm{d}^{2}\rho
}{\mathrm{d}z^{2}}\right]  =-\frac{E}{\rho}\frac{\mathrm{d}}{\mathrm{d}%
z}\left[  \frac{\mu(\rho,T)}{\rho^{2}}\frac{\mathrm{d}\rho}{\mathrm{d}%
z}\right]  . \label{D.1}%
\end{equation}
whereas the nondimensional versions of the boundary conditions (\ref{2.32}%
)--(\ref{2.33}) look the same as before. One also needs the nondimensional
form of the low-density asymptotics (\ref{2.10}) of $p$ and $G$,%
\begin{equation}
p\sim T\rho,\qquad G\sim T\ln\rho\qquad\text{as}\qquad\rho\rightarrow0,
\label{D.2}%
\end{equation}
The numerics suggest that the problem involves two asymptotic zones: the
\emph{interfacial region} and \emph{van der Waals layer}; the former is near
equilibrium and the latter, out of equilibrium.

\subsection{The interfacial region\label{appD.1}}

Since the interfacial region is near equilibrium, boundary condition
(\ref{2.32}) can be rewritten in the form%
\begin{equation}
\rho\rightarrow\rho^{(l.sat)}\qquad\text{as}\qquad z\rightarrow-\infty,
\label{D.3}%
\end{equation}
and one can also omit from Eq. (\ref{D.1}) the term involving $E$. Integrating
the resulting equation and fixing the constant of integration via (\ref{D.3}),
one obtains%
\begin{equation}
G(\rho,T)-\frac{\mathrm{d}^{2}\rho}{\mathrm{d}z^{2}}=G(\rho^{(l.sat)},T).
\label{D.4}%
\end{equation}
Next, multiply (\ref{D.4}) by $\mathrm{d}\rho/\mathrm{d}z$, and integrate
again. The integral involving $G$ can be evaluated using the equality%
\[
\frac{\partial\left(  \rho G-p\right)  }{\partial\rho}=G
\]
[which follows from identity (\ref{2.4})], and then condition (\ref{D.3}) can
be used to fix the constant of integration. Eventually, one
obtains\begin{widetext}%
\[
\frac{1}{2}\left(  \frac{\mathrm{d}\rho}{\mathrm{d}z}\right)  ^{2}=\rho\left[
G(\rho,T)-G(\rho^{(l.sat)},T)\right]  -p(\rho,T)+p(\rho^{(l.sat)},T).
\]
Recalling the Maxwell construction (\ref{2.5})--(\ref{2.6}), one can replace
in this equality%
\[
G(\rho^{(l.sat)},T)\rightarrow G(\rho^{(v.sat)},T),\qquad p(\rho
^{(l.sat)},T)\rightarrow p(\rho^{(v.sat)},T),
\]
and then use the low-density asymptotics (\ref{D.2}) to obtain%
\begin{equation}
\frac{1}{2}q^{2}\sim T\rho\left(  \ln\frac{\rho}{\rho^{(v.sat)}}+\frac
{\rho^{(v.sat)}}{\rho}-1\right)  \qquad\text{as}\qquad\rho,\rho^{(v.sat)}%
\rightarrow0,\label{D.5}%
\end{equation}
\end{widetext}where%
\begin{equation}
q=\frac{\mathrm{d}\rho}{\mathrm{d}\xi}. \label{D.6}%
\end{equation}
Asymptotics (\ref{D.5})--(\ref{D.6}) will be used to match the interfacial
region to the vdW layer.

\subsection{The vdW layer\label{appD.2}}

In the vdW layer, $\rho$ is small, so one can replace $p$ and $G$ with their
low-density expressions (\ref{D.2}). One can also assume%
\begin{equation}
\mu\rightarrow1\qquad\text{as}\qquad\rho\rightarrow0, \label{D.7}%
\end{equation}
which implies that the scale $\bar{\mu}$ used for nondimensionalization is
that of the low-density vapor, i.e., $\bar{\mu}=\mu_{0}$.

Thus, in the vdW layer, Eq. (\ref{D.1}) becomes%
\begin{equation}
\frac{T}{\rho}\frac{\mathrm{d}\rho}{\mathrm{d}z}-\frac{\mathrm{d}^{2}\rho
}{\mathrm{d}z^{2}}=-\frac{E}{\rho}\frac{\mathrm{d}}{\mathrm{d}z}\left(
\frac{1}{\rho^{2}}\frac{\mathrm{d}\rho}{\mathrm{d}z}\right)  . \label{D.8}%
\end{equation}
This equations is to be solved with boundary condition (\ref{2.33}).

To simplify Eq. (\ref{D.8}), multiply it by $\rho$, integrate, use
(\ref{2.33}) to fix the constant of integration, and change the variables
$\left(  z,\rho\right)  \rightarrow\left(  \rho,q\right)  $, where $q$ is
given by (\ref{D.6}). The resulting equation can be written in the form%
\begin{equation}
\frac{\mathrm{d}}{\mathrm{d}\rho}\left(  \frac{q^{2}}{2\rho}\right)  =T\left(
\frac{1}{\rho}-\frac{\rho^{(v)}}{\rho^{2}}\right)  +\frac{E}{\rho^{4}}q,
\label{D.9}%
\end{equation}
whereas boundary condition (\ref{2.33}) becomes%
\begin{equation}
q=0\qquad\text{at}\qquad\rho=\rho^{(v)}. \label{D.10}%
\end{equation}
One can readily verify that the solution of Eq. (\ref{D.9}) admits the
following asymptotics:%
\begin{equation}
\frac{q^{2}}{2}\sim\rho T\left(  \ln\rho+C\right)  +T\rho^{(v)}\qquad
\text{as}\qquad\rho\rightarrow\infty, \label{D.11}%
\end{equation}
where $C$ is an undetermined constant. It can be fixed by matching the inner
solution (\ref{D.11}) to asymptotics (\ref{D.5}) of the outer solution.

\subsection{Matching\label{appD.3}}

The applicability region of the outer (interfacial) solution and that of the
inner (vdW layer) solution overlap if
\[
\rho^{(v.sat)}\ll\rho\ll\rho^{(l.sat)}.
\]
In this region, Eq. (\ref{D.5}) (the inner expansion of the outer solution)
should match Eq. (\ref{D.11}) (the outer expansion of the inner solution),
which yields%
\begin{equation}
C=-1-\ln\rho^{(v.sat)}. \label{D.12}%
\end{equation}
Boundary-value problem (\ref{D.9})--(\ref{D.12}) determines the function
$q(\rho)$ and, more importantly, the dependence of the evaporation rate $E$ on
the temperature $T$ and the relative humidity $H=\rho^{(v)}/\rho^{(v.sat)}$.

$T$ and $H$ can actually be separated by representing $E$ in the form%
\begin{equation}
E=T^{1/2}\rho^{(v.sat)5/2}\,\tilde{E}_{D}(H). \label{D.13}%
\end{equation}
To find the function $\tilde{E}(H)$, substitute expression (\ref{D.13}) into
boundary-value problem (\ref{D.9})--(\ref{D.12}) and carry out the following
change of variables:%
\begin{equation}
\rho=\rho^{(v.sat)}\tilde{\rho},\qquad q=\left(  T\rho^{(v.sat)}\right)
^{1/2}\,\tilde{q}, \label{D.14}%
\end{equation}
which yields%
\begin{equation}
\frac{\mathrm{d}}{\mathrm{d}\hat{\rho}}\left(  \frac{\tilde{q}^{2}}%
{2\tilde{\rho}}\right)  =\frac{1}{\tilde{\rho}}-\frac{H}{\tilde{\rho}^{2}%
}+\frac{\tilde{E}_{D}}{\tilde{\rho}^{4}}\tilde{q}, \label{D.15}%
\end{equation}%
\begin{equation}
\tilde{q}=0\qquad\text{at}\qquad\tilde{\rho}=H, \label{D.16}%
\end{equation}%
\begin{equation}
\frac{\tilde{q}^{2}}{2}\sim\tilde{\rho}\left(  \ln\tilde{\rho}-1\right)
+H\qquad\text{as}\qquad\tilde{\rho}\rightarrow\infty. \label{D.17}%
\end{equation}
Evidently, equation (\ref{D.15}) and boundary conditions (\ref{D.16}%
)--(\ref{D.17}) involve neither $T$ nor $\rho^{(v.sat)}$ -- hence, $\tilde
{E}_{D}$ depends only on $H$, as required.

To find the spatial scale of the vdW layer, recall that, nondimensionally, it
equals the ratio of the scales of $\rho$ and $q$ in scaling (\ref{D.14}).
Recalling also that the spatial variable has been nondimensionalized on
$l_{F}$, one recovers estimate (\ref{3.8}).

Eqs. (\ref{3.6})--(\ref{3.7}) of the main body of the paper can be recovered
by re-dimensionalizing expression (\ref{D.13}).

\section{Evaporation of a liquid into its vapor: the $H\rightarrow1$ limit of
the Vlasov model\label{appE}}

Let%
\begin{equation}
\varepsilon=1-H. \label{E.1}%
\end{equation}
When expanding boundary-value problem (\ref{2.32})--(\ref{2.33}), (\ref{3.2})
in $\varepsilon$, one can show that the leading order is described by the
equilibrium solution $\rho^{(sat)}(z)$, and the first-order solution exists
subject to a certain orthogonality condition which determines $E$.

The calculation outlined above is straightforward but cumbersome. One can
by-pass it by deriving the orthogonality condition directly from the exact
boundary-value problem. It should satisfy two requirements:

\begin{enumerate}
\item[(i)] If expanded in $\varepsilon$, the zeroth order of this condition
should cancel out.

\item[(ii)] The first order should not involve the (unknown) density
$\rho^{(l)}$ of the liquid, so that $E$ is the \emph{only} unknown.
\end{enumerate}

\noindent This shortcut was used in Ref. \cite{Benilov22a} for the DIM -- and
it can be used, in exactly the same form, for the Vlasov model.

Consider the following combination of Eq. (\ref{3.2}) and isobaricity
condition (\ref{2.36}):%
\[
\int\left(  \ref{3.2}\right)  \mathrm{d}z+\frac{1}{\rho^{(l)}}\times\left(
\ref{2.36}\right)  .
\]
After straightforward algebra [involving integration by parts of the
right-hand side of (\ref{3.2}) and use of boundary conditions (\ref{2.32}%
)--(\ref{2.33})], one obtains\begin{widetext}%
\[
\hat{G}(\rho^{(v)},T)-\rho^{(v)}\int\Psi(z-z^{\prime})\,\mathrm{d}z^{\prime
}-\hat{G}(\rho^{(l)},T)+\rho^{(l)}\int\Psi(z-z^{\prime})\,\mathrm{d}z^{\prime
}-\frac{p(\rho^{(v)},T)-p(\rho^{(l)},T)}{\rho^{(l)}}=-E\int\frac{\mu(\rho
,T)}{\rho^{4}}\left(  \frac{\mathrm{d}\rho}{\mathrm{d}z}\right)
^{2}\mathrm{d}z.
\]
Recalling Eqs. (\ref{2.20}) and (\ref{2.12}), one can express the thermal
chemical potential $\hat{G}$ through its full counterpart $G$ to obtain%
\begin{equation}
G(\rho^{(v)},T)-G(\rho^{(l)},T)-\frac{p(\rho^{(v)},T)-p(\rho^{(l)},T)}%
{\rho^{(l)}}=-E\int\frac{\mu(\rho,T)}{\rho^{4}}\left(  \frac{\mathrm{d}\rho
}{\mathrm{d}z}\right)  ^{2}\mathrm{d}z.\label{E.2}%
\end{equation}
This (exact) equality can be simplified asymptotically using the fact that the
vapor is nearly saturated,%
\[
\rho^{(v)}=\left(  1-\varepsilon\right)  \rho^{(v.sat)},
\]
and the solution is close to equilibrium -- i.e.,%
\[
\rho=\rho^{(sat)}(z)+\mathcal{O}(\varepsilon),\qquad E=\mathcal{O}%
(\varepsilon),\qquad\rho^{(l)}=\rho^{(l.sat)}+\mathcal{O}(\varepsilon).
\]
Expanding Eq. (\ref{E.2}) in $\varepsilon$, using the Maxwell construction
(\ref{2.5})--(\ref{2.6}) to ascertain that the zeroth order cancels out, and
using identity (\ref{2.4}) to simplify the first order, one obtains%
\begin{equation}
\varepsilon\left(  \frac{\rho^{(v.sat)}}{\rho^{(l.sat)}}-1\right)  \left[
\frac{\partial p(\rho,T)}{\partial\rho}\right]  _{\rho=\rho^{(v.sat)}%
}+\mathcal{O}(\varepsilon^{2})=-EA=\int\frac{\mu(\rho^{(sat)},T)}%
{\rho^{(sat)4}}\left(  \frac{\mathrm{d}\rho^{(sat)}}{\mathrm{d}z}\right)
^{2}\mathrm{d}z+\mathcal{O}(\varepsilon^{2}),\label{E.3}%
\end{equation}
\end{widetext}This is the desired condition which determines $E$ through the
characteristics of the saturated interface and $\varepsilon$ (the deviation of
the relative humidity from unity).

To simplify Eq. (\ref{E.3}), assume that $\rho^{(l.sat)}\gg\rho^{(v.sat)}$ --
so that the low-density asymptotics (\ref{D.2}) holds for $p$. Observe also
that the largest contribution to the integral on the right-hand side of
(\ref{E.3}) comes from the region where $\rho^{(sat)}(z)$ is small -- hence,
in this region, $\mu$ can be replaced with its low-density approximation
(\ref{D.7}).

Taking advantage of all these approximations, omitting the $\mathcal{O}%
(\varepsilon^{2})$ terms, and recalling definition (\ref{E.1}) of
$\varepsilon$, one can rewrite (\ref{E.3}) in the form%
\begin{equation}
E=\frac{T}{A}\left(  1-H\right)  , \label{E.4}%
\end{equation}
where%
\begin{equation}
A=\int\frac{1}{\rho^{(sat)4}}\left(  \frac{\mathrm{d}\rho^{(sat)}}%
{\mathrm{d}z}\right)  ^{2}\mathrm{d}z. \label{E.5}%
\end{equation}
The coefficient $A$ has arisen before in Refs.
\cite{Benilov20d,Benilov22a,Benilov22b,Benilov23c,Benilov23d} where
evaporation has been examined using the DIM. The present results suggest that
$A$ arises in \emph{all} hydrodynamic models involving evaporation and vdW force.

Eq. (\ref{3.17}) of the main body of the paper can be recovered by
re-dimensionalizing expression (\ref{E.5}).

\section{Proof of property (\ref{A.6})\label{appF}}

Note that Eqs. (\ref{2.18}) and (\ref{2.20}) imply that%
\begin{equation}
2\pi\int_{0}^{\infty}\int_{0}^{\infty}\Phi\left(  \sqrt{z^{2}+r_{\bot}^{2}%
}\right)  \,r_{\bot}\mathrm{d}r_{\bot}\mathrm{d}z=a, \label{F.1}%
\end{equation}
Next, consider definition (\ref{A.5}) of $\Omega(r,r^{\prime})$ in the limit%
\begin{equation}
r,r^{\prime}\rightarrow\infty,\qquad r-r_{1}=\mathcal{O}(1), \label{F.2}%
\end{equation}
and observe that the largest contribution to the integral on the right-hand
side of (\ref{F.1}) comes from the region $\alpha\rightarrow0$. Expanding,
thus, the integrand of (\ref{A.5}) in $\alpha$ and introducing $r_{\bot}%
=\sqrt{rr^{\prime}}\alpha$, on obtains%
\[
\Omega(r,r^{\prime})\,rr^{\prime}\sim2\pi\int_{0}^{\infty}\Phi\left(
\sqrt{\left(  r-r^{\prime}\right)  ^{2}+r_{\bot}^{2}}\right)  r_{\bot
}\mathrm{d}r_{\bot}.
\]
Under the limit (\ref{F.2}), one can replace in the above expression
$rr^{\prime}$ with $r^{\prime2}$. Integrating the resulting equality with
respect to $r^{\prime}$ from $r$ to $\infty$, changing the variable of
integration on the right-hand side from $r^{\prime}$ to $z=r^{\prime}-r$, and
recalling Eq. (\ref{F.1}), one obtains property (\ref{A.6}) as required.

\bibliography{}

\end{document}